\documentclass{article}%
\usepackage{amsmath}
\usepackage{amsfonts}
\usepackage{amssymb}
\usepackage{graphicx}%
\begin{document}
{\huge Quantum signatures of \ }

{\huge Solar System dynamics}

$\ \ \ \ \ \ \ \ \ \ \ \ \ \ \ \ \ \ \ \ \ \ \ \ \ \ \ \ \ \ \ \ \ $

\ \ \ \ \ \ \ \ \ \ \ \ \ \ \ \ \ \ \ \ \ \textbf{Arkady L. Kholodenko}

\bigskip\ \ \ \ \ \ \ \ \ \ \ \ \ \ \ \ \ \ \ \ \ \ \ \ \ \ \ \ \ \ \ \ \ \ \ \ \ \ \ \ \ \ \ \ \ \ \ \ \ \ \ \ \ \ \ \ \ \ \ \ \ \ \ \ \ \ \ \ \ \ \ \ \ \ \ \ \ \ \ \ \ \ \ \ \ \ \ \ \ \ \ \ \ \ 

\textbf{Abstract} Let\ $\omega(i)$ be period of rotation of\ the $i$-th planet
around the Sun (or $\omega_{j}(i)$ be period of rotation of\ $j$-th satellite
around the $i$-th planet). From empirical observations it is known that within
margins of experimental errors $\sum\nolimits_{i}n_{i}\omega(i)=0$\ (or
$\sum\nolimits_{j}n_{j}\omega_{j}(i)=0)$ for some integers\ $n_{i}$
(or\ $n_{j}$ ), different for different satellite\ systems. These conditions,
known as resonance\ conditions, make uses of \ theories such as KAM\ difficult
to implement. The resonances in Solar System are similar to those encountered
in old quantum mechanics where applications of methods of celestial mechanics
to atomic and molecular physics were highly successful. With such a successes,
the birth of new quantum mechanics is difficult to understand. In short, the
rationale for its birth lies in simplicity with which the same type of
calculations can be done using methods of quantum mechanics capable of taking
care of resonances. The solution of quantization puzzle was found by
Heisenberg. In this paper \ new uses of Heisenberg's ideas are found. When
superimposed with the equivalence principle of general relativity, they lead
to quantum mechanical treatment of observed resonances in the Solar System. To
test correctness of theoretical predictions the number of allowed stable
orbits for planets and for equatorial \ stable orbits of satellites of heavy
planets is calculated resulting in good agreement with observational
data.\ \ In addition, the paper briefly discusses quantum mechanical nature of
rings of heavy planets and potential usefulness of the obtained results for cosmology.\ \ \ \ \ \ \ \ \ 

\ \ \ \ \ \ \ \ \ \ \ \ \ \ \ \ \ \ \ \ \ \ \ \ \ \ \ \ \ \ \ \ \ \ \ \ \ \ \ \ \ \ \ \ \ \ \ \ \ \ \ \ \ \ \ \ \ \ \ \ \ \ \ \ \ \ \ \ \ \ \ \ \ \ \ \ \ \ \ \ \ \ \ \ \ \ \ \ \ \ \ \ \ \ \ \ \ \ \ \ \ \ \ \ \ \ \ \ \ \ \ \ \ \ \ \ \ \ \ \ \ \ \ \ \ \ \ \ \ \ \ \ \ \ \ \ \ \ \ \ \ \ \ \ \ 


\bigskip

-----------------------------------

A.L.Kholodenko (mail) 

375 H.L.Hunter Laboratories, Clemson\textit{\ }University, Clemson, SC
29634-0973, 

USA. e-mail: string@clemson.edu

\pagebreak

\section{Introduction}

\subsection{General comments}

The role of celestial mechanics in development of modern quantum mechanics is
well described in lecture notes by Born [1] . Surprisingly, usefulness of the
atomic mechanics to problems of celestial mechanics has been recognized only
very recently [2,3]. Closely related to\ these papers is the paper by Convay
at al [4] where methods of optimal control and genetic algorithms were used
for mission planning problems\footnote{That is problems involving optimal
interplanetary travel in Solar System.}. In this work we extend the emerging
reverse trend. For this purpose it was nesessary to critically reanalyzed the
logical steps leading from the old to new quantum mechanics in the light of
available astronomical observational data. For the sake of uninterrupted
reading, some not widely known facts from history of quantum mechanics are
presented in nontraditional \ setting involving the latest results from
mathematical and atomic physics. These facts are very helful for formulation
of the problems to be solved in this paper. Thus, below, we discuss some
historical background information first.

\subsection{ Resonances in old atomic mechanics}

In 1923-24 academic year in G\"{o}ttingen Max Born replaced planned \ two-
semester lecture course in celestial mechanics by the course in atomic
mechanics. Contrary to the standard superficial descriptions of "old" quantum
mechanics which can be found at the beginning of any textbook on quantum
mechanics, the achievements of "old" quantum mechanics go far beyond
calculation of spectra of Hydrogen atom. In fact, the optical and X-ray
spectra of almost all known at that time elements were found accounting even
for the fine structure relativistic effects. The theory of quantum angular
momenta was developed and used in the theory of polyatomic molecules. The
effects of Zeemann and Stark were considered as well, etc. If one would make
an itemized list of problems considered in "old" quantum theory and \ would
compare it with that for "new" quantum theory, surprisingly, one would not be
able to find an item which was \textbf{not} treated within the "old"
formalism. \ With such an impressive list of acomplishments it is hard to
understand why this formalism was abruptly abandoned in favour of "new"
quantum mechanics in 1925. To explain this, we would like to bring some
excerpts from the paper by Pauli and Born [5]. Being thoroughly familiar with
works by Poincare$^{\prime}$ on celestial mechanics, they were trying to apply
these methods to multielectron atoms. For this purpose they were using methods
of theory of perturbations to account for electron-electron interactions. \ By
doing so they obtained the same types of divergencies as were known already
from calculations of planetary dynamics. By realizing the asymptotic nature of
the obtained expressions, they decided \ that to keep just few terms in these
expansions is the best way to proceed. By doing so a reasonably good agreement
with experimentally known location of spectral lines was expected to be
obtained. Such a state of affairs had caused frustration for Bohr who conceded
that \textsl{only those dynamical systems which admit a complete separation of
variables are quantizable}$\footnote{E.g. read [5].}$. If such a separation is
absent, according to Bohr's current opinion, the system should not possess a
discrete spectrum so that visible lines in spectra of elements other than
Hydrogen should/must be much wider. On the theoretical side such an assumption
calls for development of methods enabling to determine the widths of spectral
lines and of distribution of the intensity within these widths. Such an
intensity is expected to be connected with the underlying mechanical motion
inside the atomic system.

Spectroscopical data for almost entire periodic system were readily available
at the turn of the 20th century [6]. Bohr was well aware of these data and
used them for his search for correct atomic model (along with Rutherford's
results of 1912 on scattering from the Hydrogen atom). In particular, he
looked at the data for Helium in 1913 and published his findings in Nature
[7]. For the sake of arguments which will follow, we found it helpful to
reproduce some of the data from\ his Table 1 below.

\bigskip

\ \ \ \ \ \ \ \ \ \ \ \ \ \ \ \ \ \ \ \ \ \ \ \ \ \ \ \ \ \ \ \ \ \ \ \ \ \ \ \ \ \ \ \ \ \ \ Table
1 \
\[%
\begin{tabular}
[c]{|c|c|c|c|c|}\hline
Spectral series & $\lambda\cdot10^{8}$ & \%error & $\lambda(\frac{1}{n_{1}%
^{2}}-\frac{1}{n_{2}^{2}})\cdot10^{10}$ & $(n_{1};n_{2})$\\\hline
P$_{1}$ & 4685.98 & 0.01 & 22779.1 & (3;4)\\\hline
P$_{2}$ & 3203.30 & 0.05 & 22779.0 & (3;5)\\\hline
P$_{2}$ & 2306.20 & 0.10 & 22777.3 & (3;9)\\\hline
P$_{1}$ & 2252.88 & 0.10 & 22779.1 & (3;10)\\\hline
S & 5410.5 & 1.0 & 22774 & (4;7)\\\hline
S & 4541.3 & 0.25 & 22777 & (4:9)\\\hline
S & 4200.3 & 0.5 & 22781 & (4;11)\\\hline
\end{tabular}
\]
These data were compared with those for the Hydrogen for which he used the
analogous table (Table 2)\footnote{Which we do not reproduce.}. \ For some
reason, the data in his Table 2 \textsl{did not} contain the error column.
Since the wavelength $\lambda$ in both cases was measurable, it was possible
to evaluate the ratio K$_{H}/$K$_{He},$ where K$=\lambda(\frac{1}{n_{1}^{2}%
}-\frac{1}{n_{2}^{2}})\cdot10^{10},$ which was found to be 4.0016. At the same
time, Bohr's own calculations gave for K the following value: K$=\frac
{c(M+m)h^{3}}{2\pi^{2}Z^{2}e^{2}Mm},$ with $h$ being the Planck's constant,
$Z$ and $M$ being the charge and the mass of the nucleus, $c$ being the speed
of light and $e$ and $m$ are being the charge and the mass of the electron. By
assuming $M_{He}=4M_{H}$ and $Z_{He}=2Z_{H}$ , one readily obtains for
K$_{H}/$K$_{He}$ the result: 4.00163. It is in good agreement with that
obtained experimentally. In doing such calculations Bohr assumed that each
electron in Helium can be treated as if it is a Hydrogen-like. This surely
implies that the width of spectral lines for Helium should be practically the
same as those for the Hydrogen atom. Nowadays we know [8] that all atomic
spectra have some finite linewidth. This linewidth is determined by factors
such as: a) the collisional broadening, b) the Doppler broadening and c) the
natural broadening. Each of these is having some further ramifications. Hence,
from the standpoint of modern knowledge one can interpret Bohr's conclusions
made in 1922 as acknowledgement of the fact that spectra of elements
\textsl{other} than Hydrogen are broader because of natural reasons (so that
one should take into consideration the data from the error column in Table 1)
without invalidation of Bohr's major quantization assumptions. It should be
noted though that such a conclusion leads to the question: why the very same
factors are affecting the Hydrogen atom much less?

Unhappy with his conclusions, Bohr asked Born and Heisenberg to make more
rigorous calculations for Helium using perturbational methods analogous to
those developed in the paper by Pauli and Born in 1922$.$ Their findings were
published in 1923 and resulted in practically total failure in accurate
determination of energies of the ground and excited states for Helium atom.
This fact is documented in Born's lecture notes [1].

Since the Helium atom calculations were made by Heisenberg (under Born's
supervision) it might be not too surprising that, after all, it was Heisenberg
who found the way out of the existing difficulties. The logic of his
reasonings is discussed from the modern mathematical point of view in Section
2. \ In this introductory section we would like only to put \ his work in some
historical perspective. For this, we need to make few comments on his joint
work with Born. For the sake of space, we refer our readers to the cited
literature for details.

\subsubsection{3-body problem and the He atom (old quantum mechanics results)}

The unperturbed Hamiltonian for He was chosen as $H=$-$A$(J$_{1}^{-2}+$%
J$_{2}^{-2})$ with the constant $A=2\pi^{2}e^{4}mZ^{2}$ while the perturbation
was chosen as $H_{1}=e^{2}/R$ \ with J$_{1\text{ }}$and J$_{2\text{ }}$ being
the Bohr-Sommerfeld (B-S)\ adiabatic action integrals for electron 1 and 2 and
$R$ being the Euclidean distance between them. After H, He is the first
nontrivial 3-body mechanical system whose behavior is amended \ in accordance
with the rules of old quantum mechanics by requiring both of these integrals
(i.e. J$_{1\text{ }}$and J$_{2\text{ }})$ to have their lowest value, i.e.
$h,$ so that the \textsl{unperturbed} energy for He is twice that for H. Since
the energy $W$ for H is known to be $W=-\frac{A}{\text{J}^{2}},$ the frequency
$\omega$ of rotation of the electron at its stationary orbit \ in the
action-angle formulation of classical mechanics is obtained as $\omega
=\frac{\partial W}{\partial\text{J}}\sim n^{-3}$ \footnote{E.g. see page 140
of Ref.[1].}. In obtaining this result it was assumed that J is
\textsl{continuous} variable and, only after the differentiation is done, J is
assumed to be discrete: J$=nh$. It is important to realize at this point
\ that exactly \textsl{the same logic} was used in Heisenberg's paper on
quantum mechanics to be discussed in Section 2\footnote{E.g.see equation.(19)
below.}. Hence, for Helium atom \ within the approximations made the rotation
frequencies of both electrons are the same. This fact is known in mechanics
literature as \textsl{accidental degeneration}. In view of its crucial
importance for this paper, we would like to pause now in order to provide more
accurate definitions.

In terms of the action -angle (\textbf{I},$\mathbf{\varphi})$ variables the
Hamilton's equations of motion for a completely integrable system can be
written as:
\begin{equation}%
\begin{array}
[c]{cc}%
\dfrac{d\mathbf{I}}{dt}=0, & \dfrac{d\boldsymbol{\varphi}}{dt}=\dfrac{\partial
H}{\partial\mathbf{I}}\equiv\mathbf{\omega}(\mathbf{I}),
\end{array}
\tag{1}%
\end{equation}
where the boldface indicates that the dynamical system with Hamiltonian $H$ is
multicomponent (in general case). Solutions of the system (1) are: I$_{i}$
=$c_{i},\varphi_{i}=\omega_{i}(\mathbf{I})t+C_{i}$ , $i=1-N$. It is assumed
that $c_{i}$ and $C_{i}$ are some known constants. In view of this result, any
mechanical observable $\mathcal{F}(\mathbf{p},\mathbf{q})$ made of generalized
momenta $\mathbf{p}$ and generalized coordinate $\mathbf{q}$ can be Fourier
decomposed as
\begin{equation}
\mathcal{F=}%
{\textstyle\sum\limits_{\mathbf{n}=-\infty}^{\infty}}
A_{\mathbf{n}}\exp(i\mathbf{n}\cdot\mathbf{\varphi}\boldsymbol{),} \tag{2}%
\end{equation}
where $\mathbf{n}=\{n_{1},...,n_{N}\}.$ Accordingly, $\mathbf{n}%
\cdot\mathbf{\varphi=}%
{\textstyle\sum\nolimits_{i=1}^{N}}
n_{i}\varphi_{i}$. Such a Fourier decomposition is expected to exist even for
those \textsl{perturbed} systems for which the \textsl{empirically observed}
orbits are closed (as seen in the case of planets in our Solar System).
Dynamical system is considered to be \textsl{accidentally degenerate} if the
relation
\begin{equation}%
{\textstyle\sum\nolimits_{i=1}^{N}}
n_{i}\omega_{i}(\mathbf{I})=0 \tag{3}%
\end{equation}
holds for some \textsl{fixed} set of integers \textbf{n} and is
\textsl{degenerate} if (3) holds for \textsl{any} set of \textbf{n}'s. \ In
this work this condition will be alternatively called as \textsl{resonance}
condition in accord with modern terminology.

\subsubsection{3-body problem and the He atom (modern perspective)}

Failure of methods of celestial mechanics (modified by the B-S quantization
rule) to accurately compute the ground and excited states for He has led
Heisenberg to discovery of his matrix quantum mechanics in 1925. The problem
appeared to be completely solved until 1960 when some difficulties emerged
when the standard Hartree-Fock type variational calculations become inadequate
for description of doubly excited electron states [9]. The same authors notice
that by 1990 the improvements which were made in 60ies failed again,
especially for the extreme excitation regime which cannot be described using
single electron quantum numbers. The way out of existing difficulties was
associated with accurately designed semiclassical methods. The backbone of
these semiclassical descriptions "are the periodic orbits of the full
\textsl{classical} two-electron system without any approximations". Tanner at
all [9] notice (e.g. read page 523) that "The classical two-electron atom is
neither integrable nor fully chaotic. The apparently regular spectrum as well
as the breakdown of approximate quantum numbers for highly doubly excited
states and the enormous variation in the decay widths for resonances can be
understood \textsl{by studyng classical} \textsl{mechanics in detail}.
\textsl{Qualitative} results can be obtained by exploiting semiclassical
periodic orbit theory." In other words, use of classical mechanics is quite
sufficient for determination of both the ground and excited states of He and
only for extreme case of highly doubly excited He the description becomes
qualitative. Classical dynamics of He used for calculations of spectra is
essentially the dynamics of the restricted 3-body problem\footnote{Modified by
the fact that in the atomic case electrons repel each other and have the same
mass.} and, as such, exhibits chaotic and regular regimes. \ Since the
experimental information which can be deduced \ from the low lying spectral
excitations of He does not allow to disentangle regular and chaotic parts of
dynamics, it has become possible to simplify things further by recalculating
spectra of He, Li, Be, and diatomic molecules made out of these and other
atoms, and also of H$_{2}$, by revisiting Bohr's 1913 calculations [10-12].
These calculations involve a simple minded minimization of classical
functionals of the type considered by Bohr in 1913. Mathematical justification
of such a procedure was found by Chen et al [13] The accuracy of results
obtained with help of \ such classical calculations (employing however the B-S
quantization rule !) compares well with incomparably more elaborate
traditional quantum mechanical calculations. It should be noted though that in
his Nobel prize winning address Bohr (1923) was \ talking about his great
success in calculating spectra of almost all elements of the periodic table
using the B-S quantization rule and simple minimization procedure. It took
another 80 years or so to bring these calculations to the level comparable
with the best known quantum mechanical calculations!

\subsection{Resonances in celestial mechanics}

On page 265 \ of his lecture notes [1] Born writes$:$ $"$Accidental
degeneration is a rare and remarkable exception in astronomy; the odds against
(Eq.(3)) being exactly fulfilled are \textsl{infinite}. A close approach to it
is found in the case of perturbations of some minor planets (Achilles,
Patroclus, Hector, Nestor) which have very nearly the same period of
revolution as Jupiter. \textsl{In atomic theory}, on the other hand, where
J$_{k}$'s can have only discrete values, \textsl{accidental degeneration is
very common}." \ As result of such an accidental degeneracy Heisenberg's
attempt at perturbative calculations for He failed miserably. Such a failure
caused him to reconsider the whole computational scheme resulting in an
ultimate breakthrough in 1925 leading to new quantum mechanics.

Before discussing his contributions from the modern perspective, we would like
to make few remarks regarding the accuracy of astronomical data in Born's
lectures. In 1968 Molchanov [14], \ while analyzing the astronomical data,
came to the conclusion that the accidental degeneracy for Solar (and, very
recently, Solar-like [15,16] system(s) is as common as in atomic systems. In
Table 2 (below) taken from his work (Molchanov's Table 1) we reproduce some
data taken from this reference.

\ \ \ \ \ \ \ \ \ \ \ \ \ \ \ \ \ \ \ \ \ \ \ \ \ \ \ \ \ \ \ \ \ \ \ \ \ \ Table
2%
\[%
\begin{tabular}
[c]{|c|c|c|c|c|c|c|c|c|c|c|c|c|c|}\hline
& Planet & $\omega_{i}^{O}$ & $\omega_{i}^{T}$ & $\Delta\omega/\omega$ &
n$_{1}$ & n$_{2}$ & n$_{3}$ & n$_{4}$ & n$_{5}$ & n$_{6}$ & n$_{7}$ & n$_{8}$
& n$_{9}$\\\hline
1 & Mercury & 49.22 & 49.20 & 0.0004 & 1 & -1 & -2 & -1 & 0 & 0 & 0 & 0 &
0\\\hline
2 & Venus & 19.29 & 19.26 & 0.0015 & 0 & 1 & 0 & -3 & 0 & -1 & 0 & 0 &
0\\\hline
3 & Earth & 11.862 & 11.828 & 0.0031 & 0 & 0 & 1 & -2 & 1 & -1 & 1 & 0 &
0\\\hline
4 & Mars & 6.306 & 6.287 & 0.0031 & 0 & 0 & 0 & 1 & -6 & 0 & -2 & 0 &
0\\\hline
5 & Jupiter & 1.000 & 1.000 & 0.000 & 0 & 0 & 0 & 0 & 0 & 2 & -5 & 0 &
0\\\hline
6 & Saturn & 0.4027 & 0.400 & 0.0068 & 0 & 0 & 0 & 0 & 1 & 0 & -7 & 0 &
0\\\hline
7 & Uranus & 0.14119 & 0.14286 & -0.0118 & 0 & 0 & 0 & 0 & 0 & 0 & 1 & -2 &
0\\\hline
8 & Neptune & 0.07197 & 0.07143 & 0.0075 & 0 & 0 & 0 & 0 & 0 & 0 & 1 & 0 &
-3\\\hline
9 & Pluto & 0.04750 & 0.04762 & -0.0025 & 0 & 0 & 0 & 0 & 0 & 1 & 0 & -5 &
1\\\hline
\end{tabular}
\]
For satellite systems of Jupiter, Saturn and Uranus Molchanov's paper also
contains tables similar to our Table 2. According to the book by Beletsky
[17], in view of the resonance nature of our Solar System, uses of KAM theory
\ [18] for explanation of planetary stability typically fail.

To understand the data \ in Table 2 several comments are in order. First, the
displayed frequencies are measured in the system of units in which the
Jupiter's frequency was chosen as the unit of measurement. Second, in view of
\ Eq.(3), the first row of data from Table 2 should be actually read as
$\omega_{1}-\omega_{2}-2\omega_{3}-\omega_{4}=0$. All other rows should be
treated accordingly. The theoretical frequencies $\omega_{i}^{T}$ are those
which satisfy the resonance conditions exactly while $\omega_{i}^{O}$ denote
the observed frequencies. The data for Pluto should \textsl{not} to be
considered in terms of resonances for the following reason.

Consider a scalar product $\mathbf{n}\cdot\mathbf{\varphi\equiv\Lambda}$ in
Eq.(2). This can be looked upon as representation of the vector
$\mathbf{\Lambda}$ in the coordinate basis \{$\mathbf{\varphi\}.}$ The
coordinate basis can be changed with help of some matrix \textbf{A} so that
$\mathbf{\Lambda}=\mathbf{n}\cdot\mathbf{A}\cdot\mathbf{\tilde{\varphi}}$. It
can be argued [1],[14]\ that $\ \det\mathbf{A}=1,$ so that the matrix
\textbf{A} must be a unimodular square matrix. Only for the sake of this
requirement the data for Pluto in Table 2 were assigned in a way\ given in the
Table 2. Next, $\Delta\omega/\omega$ should be understood as ($\omega_{i}%
^{O}-\omega_{i}^{T})/\omega_{i}^{O}$. After this, the obtained error margins
can be compared against those for He in Table 1. Such a comparison indicates
that the accuracy in both cases is essentially the same. It is such that Bohr
was able to obtain using his old quantum \ mechanical theory a reasonably
accurate ratio K$_{H}/$K$_{He}$ in agreement with experiment. It makes
physical sense to blame the intrinsic inaccuracy of the collected data (e.g.
that in Table 1) for observed frequency discrepancies. Hence, along with Bohr,
it is reasonable to claim that, with exception of H, other atomic systems are
not quantizable because of these discrepancies \footnote{Subsequent
developments of quantum mechanics demonstrated that Bohr was apparently wrong.
We say "apparently" in view of the results of Section 2 \ where correct
quantization prescription is discussed based on improvement of Heisenberg's
ideas. In view of results of Section 2, it is reasonable to say that Bohr's
intuition was \ nevertheless correct but the situation can be improved
rigorously using mathematical methods which were not available in Bohr's
time.}. The same reasoning should then be applied to the planetary systems,
especially in view of critique of Molchanov's work by Henon [19] and Backus
[20]. These authors argued that the error margins in Molchanov's tables are
too large for the resonances to be considered seriously.\ The comparison
between \ the Tables 1 and 2 indicates that even though the arguments by Henon
and Backus may be mathematically correct, they do not have sound physical
support due to intrinsic inaccuracies in measurements which cannot be
substantially improved\footnote{E.g.read Sections 3 and 4 where these
measurements are discussed for objects such as Solar System, etc.}. Thus,
Molchanov's data and their interpretation remain correct at the physical level
of rigor even without additional explanations made by Molchanov [21-22] in
defence of his results. Furthermore, subsequently obtained results by Brin
[23] and Patterson [24] strongly suggest that, at least pairwise, planets are
in resonance with each other. This is true in particular for the heavy planets
in our Solar System as demonstrated by Ferraz-Mello et al [16]. Evidently, the
linear combination of such pairwise resonances leads back to
the\ Molchanov-type results.

It should be noted though that \textsl{quantization prescriptions discovered
by Heisenberg remain correct even in the case when there are no resonances}:
\textsl{They} \textsl{rely only on the existence of the closed stable orbits}.
\textsl{Existence of resonances, in fact, simplifies matters considerably
since it makes the task of establishing the quantum-classical correspondence
much easier as explained in the rest of this paper.}

Prior to Molchanov's 1968 work an effort to explain the ubiquity of resonances
in Solar System using methods of classical mechanics was made by Goldreich
[25]\textbf{\ (}1965) who demonstrated that "special cases of commensurate
mean motions are not disrupted by tidal forces". Moreover, he proposed that it
is the tidal forces which drive otherwise incommensurate system to commensurability.

Thus, the problem\ of stability of our Solar System is \ very much the same as
that for the multielectron atoms. In both cases the accidental degeneracies
(resonances) preclude systematic use of standard perturbational methods.
Unlike more traditional \ classical mechanics treatments [26], we apply
Heisenberg-style arguments ultimately aimed at explanation of Solar System
stability. For the sake of space, we are not discussing \ in this work the
spin-orbit- type resonances also ubiquitous in the Solar System [26].

Finally, we would like to mention that development of our quantum mechanical
formalism proceeds in historical accord with that for atomic systems for which
the static (spectral) problems were considered first. The dynamical problems
of atoms/molecules formation as well as their stability towards disintegration
were considered only afterwards. Hence, only the spectral-type problems will
be discussed in this work.

\subsection{Organization of the rest of this paper}

Existence of \ stable closed orbits, of resonances, as well as the lack of
dissipation (in spite of presence of tidal effects) in Solar (and Solar-like)
System(s) are indicative of quantum nature of the orbital motions in the Solar
System. Nevertheless, the formalism of quantum mechanics in its traditional
form present in textbooks for students cannot be used. To apply methods of
quantum mechanics to celestial mechanics is possible with use of Heisenberg's
original ideas updated with help of the latest mathematical results. Sections
2 and 3 as well as Appendices A and B provide a self-contained overview of
quantum mechanics based on Heisenberg's ideas. They provide needed background
for the actual quantum calculations in celestial mechanics which are performed
in Section 4 (supplemented with Appendix C). The main results of this section
(and the whole paper) are summarized in Table 4. In this table the number of
stable orbits for planets of Solar System as well as the number of stable
orbits for satellites of heavy planets (Jupiter, Saturn, Uranus and Neptune)
is calculated and compared against the observed numbers. Unusually good
agreement between the calculated and observed numbers for Solar System, and
the satellite systems of Jupiter and Saturn is obtained resulting in further
suggestions for observational astronomy of the Solar System. In Table 3 a
comparative summary of main theoretical assumptions of both quantum atomic and
quantum celestial mechanics is given. Using it, the motion of rings around
heavy planets is studied briefly resulting in the same conclusions about the
quantum nature of such type of motion. Some auxiliary mathematical results
needed for these calculations are presented in Appendix D. In study of all
cases, including dynamics of planetary rings, it follows that the equivalence
principle of general relativity plays the decisive role in development of
quantum celestial mechanics. This fact caused us to write Section 5 in which
the effects of general relativity on quantum mechanics of Solar System are
studied. In it we discuss how the\ obtained results (for which the importance
of the Lorentz group SO(2,1) is emphasized) should be amended \ if we are
interested in knowing to what extent the (larger scale) symmetries of
space-time typically considered in cosmological models of general relativity
may affect the quantum dynamics of Solar System. Such an information can be
used in reverse for probing symmetries of space-time at scales comparable or
larger than that for our Solar System. Finally, Section 6 is devoted to some
concluding remarks.

\ 

\section{ Heisenberg's honeycombs and resonances}

\subsection{General comments}

In this section we discuss Heisenberg's ground breaking paper [27] on quantum
mechanics from perspective of modern mathematics. We begin with observation
that the Schr\"{o}dinger equation cannot be reduced to something else which is
related to our macroscopic experience. \textsl{It has to} \textsl{be
postulated}.\footnote{Usually used appeal to the DeBroigle wave-particle
duality is of no help since the wave function in the Schr\"{o}dinger's
equation plays an auxiliary role while the De Broigle waves are assumed to
exist in real space-time.} On the contrary, Heisenberg's basic equation from
which all quantum mechanics can be recovered is directly connected with
experimental data and looks almost trivial. Indeed, following Bohr, Heisenberg
looked at the famous equations for energy levels difference%
\begin{equation}
\omega(n,n-\alpha)=\frac{1}{\hbar}(E(n)-E(n-\alpha)), \tag{4}%
\end{equation}
where both $n$ and $n-\alpha$ are some integers. He noticed that this
definition leads to the following fundamental composition law:%
\begin{equation}
\omega(n-\beta,n-\alpha-\beta)+\omega(n,n-\beta)=\omega(n,n-\alpha-\beta).
\tag{5a}%
\end{equation}
Since by design $\omega(k,n)=-\omega(n,k),$ the above equation can be
rewritten in a symmetric form as
\begin{equation}
\omega(n,m)+\omega(m,k)+\omega(k,n)=0. \tag{5b}%
\end{equation}
In such a form it is known as the honeycomb equation (condition) in current
mathematics literature [28] where it was rediscovered totally independently of
\ Heisenberg's \ key quantum mechanical paper\ and, apparently, with different
purposes in mind. Connections between \ mathematical results of Knutson and
Tao and those of Heisenberg were discovered in the recent paper by
Kholodenko[29]. In this \ work some results of this paper will be used.

In particular, we begin by noticing that Eq.(5b) due to its purely
combinatorial origin does not contain the Plank's constant $\hbar$. Such fact
is of major importance for this work. In particular, the simplest resonance
condition encountered in celestial mechanics%
\begin{equation}
n_{1}\omega_{1}+n_{2}\omega_{2}+n_{3}\omega_{3}=0 \tag{6}%
\end{equation}
can be equivalently rewritten in the form of (5b), where $\omega
(n,m)=\omega_{n}-\omega_{m}$. It would be quite unnatural to think of the
Planck's constant for this case. Even though, the resonance condition is
equivalent to Heisenberg's quantization condition, Eq.(5b), the reverse may
not be true since frequencies in Eq.(5b) may be irrational. It should be noted
though that such irrationality would be very difficult to detect
experimentally in view of natural causes leading to the line broadening
mentioned in the Introduction. Thus, from the experimental standpoint
equations (5b) and (6) are equivalent\footnote{Although to make the
frequencies independent (and, hence, irrational) is easy mathematically, it is
unrealistic to detect such fact experimentally. In other words, even though
the critique of Molchanov's paper [14] by Henon [19]\textbf{\ } and Backus
[20] could be mathematically justified in spite of Molchanov's counter
arguments [21,22], it is only of academic value.}. \textsl{Furthermore, by
assuming irrationality we \ would run into difficulty with obtaining the
semiclassical limit in which (it is believed) the old quantum mechanics based
on the Bohr-Sommerfeld method of quantization should be applicable.These
arguments imply that, at least semiclassically, dynamics of all quantum
mechanical systems is resonant.}

Equation (5b) is the basic building block of the honeycomb structure encoding
all information about the spectra of quantum system. Details leading to
construction of this combinatorial structure \ are summarized in the paper by
Knutson and Tao [28]. They were used in Kholodenko's paper [29] in which some
physical applications absent in Knutson's-Tao paper are discussed in detail.

To describe such honeycomb structure in a nutshell, let us choose the basic Y-
shaped tripod whose edges are labeled by frequencies $\omega(n,m)^{\prime}s$
is such a way that the total sum of these labels is equal to zero, as in
Eq.(5b). The honeycomb is made of collection of such tripods placed on a 2-
dimensional plane and joined with each other in such a way that the
frequencies at the edges match. Several additional rules were set up by
Knutson and Tao and are given in their original papers [30, 31]. Our readers
encouraged at this point to consult the interactive web site designed by Tao
[32] in order to get a feeling of honeycombs as combinatorial objects. \ For
physical applications, other than those discussed in this paper, our readers
are referred to the paper by Kholodenko [29]. Provided references allow us to
squeeze to the absolute minimum the amount of \ mathematical information in
this paper.

With account of these remarks, we proceed with development of Heisenberg's
arguments. In his paper of October 7th of 1925, \ Dirac[33], \ being already
aware of Heisenberg's key paper\footnote{This paper was sent to Dirac by
Heisenberg himself.}, streamlined Heisenberg's results and \ introduced
\ notations which are in use up to this day. He noticed that the combinatorial
law, Eq.(5a), for frequencies, when used in the Fourier expansions for
composition of observables, leads to the multiplication rule
$a(nm)b(mk)=ab(nk)$ for the Fourier amplitudes for these observables. In
general, in accord with Heisenberg, one expects that $ab(nk)\neq ba(nk).$ Such
multiplication rule is typical for matrices. In the modern quantum mechanical
language such matrix elements are written as $<n\mid\hat{O}\mid m>\exp
(i\omega(n,m)t)$ so that Eq.(5b) is equivalent to the matrix statement%
\begin{align}%
{\textstyle\sum\nolimits_{m}}
&  <n\mid\hat{O}_{1}\mid m><m\mid\hat{O}_{2}\mid k>\exp(i\omega(n,m)t)\exp
(i\omega(m,k)t)\nonumber\\
&  =<n\mid\hat{O}_{1}\hat{O}_{2}\mid k>\exp(i\omega(n,k)t) \tag{7}%
\end{align}
for some operator (observables) $\hat{O}_{1}$ and $\hat{O}_{2}$ evolving
according to the rule: $\hat{O}_{k}(t)=U\hat{O}_{k}U^{-1},k=1,2,$ provided
that $U^{-1}=\exp(-i\frac{\hat{H}}{\hbar}t).$ \ From here it follows that
$U^{-1}\mid m>=\exp(-\frac{E_{m}}{\hbar}t)\mid m>$ \ if one identifies
$\hat{H}$ with the Hamiltonian operator. Clearly, upon such an identification
the Schr\"{o}dinger equation can be obtained at once as is well known [34]$.$
We shall avoid such a pathway (at least at this stage), however. Moreover, we
also shall avoid use of \ Heisenberg's equations of motion
\begin{equation}
i\hbar\frac{\partial}{\partial t}\hat{O}=[\hat{O},\hat{H}]. \tag{8}%
\end{equation}

Our readers may ask at this point: why it is necessary to do so? And, if this
is the case, what else is left from the traditional formulations of quantum
mechanics which still can be used? The answers can be found in [29,35]. For
the sake of uninterrupted reading they are summarized \ also below.

Following Heisenberg's philosophy, we shall assume that there is a set of
classical observables $\{O_{i}(t)\}$ which is assumed to be complete in the
sense that the composition of any two of these observables is given by the
classical fusion rule:
\begin{equation}
\{O_{i},O_{j}\}=%
{\textstyle\sum\nolimits_{k}}
C_{ij}^{k}O_{k}, \tag{9}%
\end{equation}
where $C_{ij}^{k}$ \ are some known constants and $\{,\}$ represents the
Poisson brackets of classical mechanics. Accordingly, quantum mechanically,
instead of \ Eq.(8), we need to consider the decomposition
\begin{equation}
\lbrack\hat{O},\hat{H}]=%
{\textstyle\sum\nolimits_{k}}
\tilde{C}_{ij}^{k}\hat{O}_{k} \tag{10}%
\end{equation}
valid for any $t$ ! Under such circumstances (quantum) dynamics formally
disappears! This is, of course, an exaggeration since not all systems possess
needed symmetry so that the fusion rule, Eq.(9), may not exist\footnote{Since
all observables are made of \textbf{p} and \textbf{q} variables, such a rule
does exist if we decompose these observables into power series in \textbf{p}
and \textbf{q}. In those cases when such a series is infinite, normally, one
should expect loss of integrability and, hence, loss of quantization. For one
dimensional \ many body systems the situation might be repairable for suitably
chosen interaction potentials. We shall elaborate on this remark further
below, in Section 2.5.}. When it does exist, such an observation can be
strengthened due to the following chain of arguments. In mathematics,
expressions like $\hat{O}_{i}(t)=\hat{U}\hat{O}_{i}\hat{U}^{-1}\equiv
Ad_{\hat{U}}\hat{O}_{i}$ define an \textsl{orbit} for the operator $\hat
{O}_{i}$ in the Lie algebra (made of operators $\{\hat{O}_{i}\})$ \ so that
the motion is caused by the action of elements $\hat{U}$ from the Lie group
associated with such an algebra. Following the existing rules and notations in
mathematics of Lie groups and Lie algebras [36], we write $ad_{\hat{H}}\hat
{O}$ \ for $[\hat{O},\hat{H}].$ This requires us to use the r.h.s. of Eq.(10)
instead of the formal symbol $i\hbar\frac{\partial}{\partial t}\hat{O}$ used
in Heisenberg's mechanics. Evidently, we can obtain the same (or even greater)
information by working with $Ad$ operators instead of $ad.$ In particular, it
is useful to consider the trace, i.e. $tr\{Ad_{\hat{U}}\hat{O}_{i}\}=\chi
(\hat{O}_{i}),$which is just the character of $\hat{O}_{i}.$ It is
time-independent by design. If there is no time evolution then, superficially,
nothing happens. This is not true, however, as was recognized long time ago by
Dirac [34]. In Chapter 9 of his book he writes : " The Hamiltonian is
symmetrical function of the dynamic variables and thus commutes with every
permutation. It follows that each permutation is a constant of motion. This
happens even if the Hamiltonian is not constant\footnote{I.e. time-dependent.}%
." At this point it is important to recall the famous theorem by Caley [37]
which states that "every finite group is isomorphic to some permutation
group". It should be noted that in mathematics literature the "permutation"
group has the same meaning as "symmetric" group $S_{n}\footnote{Here $n$
denotes the number of elements in the group.}$. In physics and, especially, in
quantum mechanics, the symmetric group can be infinite dimensional. The theory
of such groups was unknown to Dirac since it was developed only quite recently
[38]. This fact explains why it\ have not been in use in the traditionally
written textbooks on quantum mechanics. Fortunately, for the purposes of
\ this work, it is sufficient to use only a tiny fraction from the theory of
symmetric groups.

\subsection{Some useful facts about S$_{n}$}

\bigskip

In view of the fact that the character $\chi(\hat{O}_{i})=tr\{Ad_{\hat{U}}%
\hat{O}_{i}\}$ is manifestly time-independent, the orbit $Ad_{\hat{U}}\hat
{O}_{i}$ is caused by permutations\footnote{Since, according to Dirac, the
permutation operator commutes with the Hamiltonian.}. These can be analyzed
using methods of algebraic geometry [39] and theory of linear algebraic groups
[40]. A brief and self-contained introduction to these topics can be found in
Kholodenko [35]. The key concept in this field is the notion of the torus
action $T$. It is directly connected with the notion of the Weyl-Coxeter
reflection group $W=N/T$ \ in which the numerator $N$ refers to some
permutation group and the denominator $T$ refers to those group elements
(fixed points) which remain unaffected by permutations. Representations of Lie
algebras (including the affine Lie algebras) associated with these
Weyl-Coxeter reflection groups produce all \ Lie algebras known in quantum
mechanics and in conformal field theories [41]. To simplify matters, we choose
another pathway in this work to arrive at the same results. It is better
adapted for connecting the experimental data with theoretical constructions.

We begin with observation that the representation theory for $S_{n}$ can be
built using representation theory for general linear group $GL$(N,\textbf{C})
acting in the complex space made of $n$ copies of \textbf{C}$^{N},$ $i.e.$
\textbf{C}$^{N}$ $\otimes$\textbf{C}$^{N}\otimes\cdot\cdot\cdot\otimes
$\textbf{C}$^{N}$. This fact is known as the Schur-Weyl duality [42]. The
Schur functions (to be defined below) are characters of $GL$(N,\textbf{C)}%
.\textbf{\ }They play the key role in developing the representation theory of
$S_{n}$ in which both $N$ and $n$ can become infinite.

Next, we recall that a \textsl{partition} $\lambda$ (finite or infinite) is a
sequence%
\begin{equation}
\lambda=\{\lambda_{1},\lambda_{2},\lambda_{3},...\} \tag{11}%
\end{equation}
made of integers\footnote{\ Since these numbers normally are identified with
the eigenvalues of some matrix (finite or infinite) [28,29], one can relax the
condition that $\lambda_{i}^{\prime}s$ are integers and make them rational or
even irrational numbers but the nonnegativity and the ordering are essential.}
such that $\lambda_{1}\geq\lambda_{2}\geq\lambda_{3}\geq\cdot\cdot\cdot\geq0$.
The \textsl{weight} of $\lambda$ is denoted by $\left\vert \lambda\right\vert
=%
{\textstyle\sum\nolimits_{i}}
\lambda_{i}$. If $\lambda^{\prime}s $ are integers, and if $\left\vert
\lambda\right\vert =n$ we say that $\lambda$ is a \textsl{partition} of $n$.
Let $\lambda$ be a partition. It is useful to associate with it a monomial
$\mathbf{x}^{\lambda}\equiv x_{1}^{\lambda_{1}}x_{2}^{\lambda_{2}}\cdot
\cdot\cdot$ . Next, we introduce a symmetric function $m_{\lambda}$ as a sum
of all distinct monomials that can be obtained from $\mathbf{x}^{\lambda}$ by
permuting of \ all arguments. Using these results it is possible to prove [43]
that the Schur function $s_{\lambda}$ can be \ represented with help of
$m_{\lambda} $ as follows
\begin{equation}
s_{\lambda}=m_{\lambda}+%
{\textstyle\sum\limits_{\mu<\lambda}}
K_{\lambda\mu}m_{\mu}. \tag{12}%
\end{equation}
To explain the meaning of the Kostka number $K_{\lambda\mu}$ in (12) we should
mention the one-to -one correspondence between the partitions and the Young
tableaux [39]. In terms of such a correspondence the Kostka number
$K_{\lambda\mu}$ is just the number of semistandard tableaux with shape
$\lambda$ and weight $\mu.$ Hence, for not too large tableaux such a number
can be straightforwardly computed. \ The Schur functions possess a remarkable
orthogonality property. For partitions $\lambda$ and $\mu$ and properly
defined scalar product $<,>$ one can write
\begin{equation}
<s_{\lambda},s_{\mu}>=\delta_{\lambda,\mu} \tag{13}%
\end{equation}
in accord with general theory of characters and, in particular, of characters
of $S_{n}$ [43]. With such defined orthogonality property of $s_{\lambda
}^{\prime}s$ one can proceed with the composition (fusion) law for Schur
functions. It is given by
\begin{equation}
s_{\lambda}\cdot s_{\mu}=%
{\textstyle\sum\limits_{\nu}}
C_{\lambda\mu}^{\nu}s_{\nu}, \tag{14}%
\end{equation}
where $\left\vert \lambda\right\vert +\left\vert \mu\right\vert =\left\vert
\nu\right\vert $ and, in view of (13), the Littlewood-Richardson (L-R)
coefficient $C_{\lambda\mu}^{\nu}$ can be formally defined as $C_{\lambda\mu
}^{\nu}=$ $<s_{\lambda}\cdot s_{\mu},s_{\nu}>.$These coefficients play an
important role in representation theory of $S_{n}$ analogous to the role the
Clebsch-Gordan coefficients play in the \ representation theory for spin and
angular momenta. The L-R coefficients can be obtained \ very easily with help
of the honeycomb construction as discussed by Knutson and Tao [28] and by
Kholodenko [29]. \ For completeness, we provide a brief sketch of how this can
be done.

We begin with the 1-honeycomb. It is just the Y-shaped tripod as discussed
already. When constructing the 2-honeycomb in the plane we shall follow the
rule that the labels for the edges of this new honeycomb should be
geometrically and combinatorially arranged in the same way as those for the
1-honeycomb. This requires us to use yet another two tripods which can be
joined together \ and with the third tripod only in one way in view of the
imposed rules\footnote{E.g. see Fig.2 in Kholodenko's paper.[29].}. Thus,
instead of just one boundary label, e.g. $\lambda_{1},$ in the North-West
direction, now we shall have two, say, $\lambda_{1}$ and $\lambda_{2}.$ The
same applies for the South and the North-East directions. Thus, all larger
honeycombs will have only the boundaries in the directions just mentioned
which are labeled by the partitions $\lambda,\nu$ and $\mu$. Unlike the\ 2-
honeycomb for which the boundary labels determine such a honeycomb uniquely,
for larger honeycombs this is no longer true. For the fixed set of boundary
labels, normally, there will be more than one honeycomb with these labels. On
page 1053 of Knutson and Tao paper [30] the following theorem is proven: Let
$\lambda,\mu$ and $\nu$ \ be three pre assigned (boundary) partitions for the
$k$-honeycomb. Then the number of different honeycombs with such pre assigned
boundary conditions is given by the L-R coefficient $C_{\lambda\mu}^{\nu}.$

Summarizing, we have defined a set (finite or not) of \ mutually orthogonal
Schur polynomials which by design forms the Hilbert space. The partitions and
the energy levels can be put into one-to-one correspondence by using the
honeycomb condition, Eq.(5). Such Hilbert space is designed using experimental
data. We can look at different portions (segments) of the spectra and study
their overlaps thanks to the composition rule, Eq.(14)\footnote{Very much like
it is done \ in the case of determination of the entire DNA structure from its
fragments.}. Unlike more traditional \ formulations of quantum mechanics
requiring objects of classical mechanics as an input, no reference to the
objects of classical mechanics was made thus far. In the next (sub)sections we
shall discuss the extent \ to which such a way of developing quantum mechanics
is advantageous as compared with more traditional formulations.

\subsection{From combinatorics to physics}

\bigskip

In this subsection we follow logic of Heisenberg's paper once again. In this
paper Heisenberg was also concerned with proper interrelation between the
objects of classical and quantum mechanics. Naturally, he focused his
attention at the Bohr-Sommerfeld (B-S) quantization rule%
\begin{equation}%
{\textstyle\oint}
pdq=nh,\text{ \ }n=0,1,2,..., \tag{15}%
\end{equation}
since this rule was the only one available link between the new and old
mechanics. He argued that such a rule is not exact! It is determined with
accuracy up to a constant (unknown at the time of writing of his paper). If
such a constant would be known, the B-S rule would\textsl{\ }become exact,
that is valid for any $n$'s.

From the point of view of our present understanding of quantum mechanics
Heisenberg's intuition was correct: the old \ fashioned B-S rule is valid
rigorously only in the limit of large $n$'s while the calculation of the
constant can be done, for instance, with help of either the WKB or
considerably more sophisticated theory of Maslov indices [44]. As much as
these arguments are plausible, they are nevertheless superficial as can be
found from reading the page 246 of the book by Arnol'd \ [45]. Using this
reference it follows that already at the classical level the adiabatic
invariant $%
{\textstyle\oint}
pdq$ is determined only up to some constant. This observation makes
Heisenberg's arguments less convincing. Nevertheless, following Heisenberg, we
assume that if the B-S quantization rule is corrected, \ it would make sense
fully quantum mechanically. Presumably, under such circumstances one can get
an additional information out of it. For this purpose, \ Heisenberg introduces
the Fourier decomposition of the generalized coordinate $q$ as
\begin{equation}
q(n,t)=%
{\textstyle\sum\limits_{\alpha=-\infty}^{\infty}}
a_{\alpha}(n)\exp(i\omega(n,\alpha)t), \tag{16}%
\end{equation}
where,\ in anticipation of its quantum mechanical use, it is written with
respect to some pre assigned energy level $n$. Using Eq.(16) the velocity can
be readily obtained as follows%
\begin{equation}
\dot{q}(n,t)=%
{\textstyle\sum\limits_{\alpha=-\infty}^{\infty}}
ia_{\alpha}(n)\omega(n,\alpha)\exp(i\omega(n,\alpha)t). \tag{17a}%
\end{equation}
The calculation of the velocity square over the total period is given
therefore by
\begin{equation}%
{\textstyle\oint}
dt[\dot{q}(n,t)]^{2}=2\pi%
{\textstyle\sum\limits_{\alpha=-\infty}^{\infty}}
\left\vert a_{\alpha}(n)\right\vert ^{2}\omega(n,\alpha)^{2}. \tag{17b}%
\end{equation}
In view of this result, the B-S adiabatic invariant can be rewritten as
\begin{equation}%
{\textstyle\oint}
pdq=%
{\textstyle\oint}
m\dot{q}dq=%
{\textstyle\oint}
m\dot{q}^{2}dt=2\pi m%
{\textstyle\sum\limits_{\alpha=-\infty}^{\infty}}
\left\vert a_{\alpha}(n)\right\vert ^{2}\omega(n,\alpha)^{2}=nh+const.
\tag{18}%
\end{equation}
Next, Heisenberg proceeds as follows. Since the \textit{const} is unknown, it
is of interest to obtain results which are constant-independent. At the same
time, since the result, Eq.(18), is assumed to be exact, we have to use
instead of scalars $\left\vert a_{\alpha}(n)\right\vert ^{2}$ the matrices in
accord with Eq.(7). This causes us to use matrices of the type $\left\vert
a(n,n+\alpha)\right\vert ^{2}$ and $\left\vert a(n,n-\alpha)\right\vert ^{2}$
depending on the actual sign of $\alpha.$ In addition, he had silently assumed
that the $n$ -dependence for amplitudes is much weaker than that for the
frequencies $\omega(n,n+\alpha)$ and $\omega(n,n-\alpha)$ so that it can be
neglected completely. Under such conditions he treats $n$ as continuous
variable and differentiates both sides of Eq.(18) with respect to $n$ thus
obtaining the following result:

\bigskip%
\begin{equation}
h=4\pi m%
{\textstyle\sum\limits_{\alpha=0}^{\infty}}
\{\left\vert a(n,n+\alpha)\right\vert ^{2}\omega(n,n+\alpha)-\left\vert
a(n,n-\alpha)\right\vert ^{2}\omega(n,n-\alpha)\}. \tag{19}%
\end{equation}
Obtained result takes into account that $\omega(mn)=-\omega(nm).$ The validity
of this result depends upon additional assumption about the ground state
energy. If $n_{0}$ represents such a state, then one must require that
$a(n_{0},n_{0}-\alpha)=0$ for all $\alpha>0.$ When (19) is used in combination
with the results from Appendix A, the famous commutation rule
\begin{equation}
\lbrack\hat{x},\hat{p}]=i\hbar\tag{20}%
\end{equation}
is obtained. From the above derivation and results of Appendix A several
conclusions can be drawn.

First, the number of x-p commutators by construction is in one-to one
correspondence with the number of the B-S adiabatic invariants. This means
that the system which is completely integrable classically can be completely
quantized. However, if classically system is nonintegrable, one cannot write
the classical Hamiltonian and to replace x's and p's in it by the
corresponding operators obeying commutation relations, Eq.(20), for each
generalized degree of freedom. Although this prescription is used routinely in
the existing textbooks on quantum mechanics, rigorously speaking, one cannot
write down the Schr\"{o}dinger's equation in such a case so that formally
Bohr's intuition was correct.

Second, Eq.(19) assumes\ that the underlying mechanical system, \ when \ it is
written in terms of the action-angle variables, is essentially the set of
independent harmonic oscillators. Heisenberg's derivation explicitly assumes
that quantum mechanically there is a ground state-typical for the harmonic
oscillator- but otherwise the spectrum is boundless. If the system is
nonintegrable, again, the commutation rule, Eq.(20), is not justified. Hence,
once again, one cannot write the Schr\"{o}dinger's equation. To by pass this
difficulty Heisenberg developed perturbation theory (for one dimensional case
only!) which uses the classical perturbation theory as an input modified by
the imposed (quantum) requirements on amplitudes and frequencies. Although he
did not discuss the resonances, he did checked that the obtained perturbative
expressions for energies are in agreement with the basic equations (4) and (5)
assuring correct quantization. No attempts to study the multidimensional case
was made.

Third, as results of Appendix A demonstrate, the experimental justification of
the commutator rule, Eq.(20), is based on the validity of results of the first
order perturbational calculations. Mathematically, such a procedure is
questionable or, better, might be totally unacceptable.

Furthermore, the B-S quantization cannot be used for spin quantization (since
formally there is no classical analog of spin, i.e. the B-S rule does not
account for half integers). The spin has no place in the Schr\"{o}dinger
formalism, and, \textsl{apparently}, there is no room for spin in Heisenberg's
formalism as well. Fortunately, this happens only apparently as we would like
to discuss now. This is possible only because the facts just listed \ do not
affect the main Heisenberg's quantization postulate, Eq.(5), which, as
recognized already by Heisenberg, is \textsl{more fundamental} than the x-p
commutator identity.

\subsection{From physics back to combinatorics}

\bigskip

To find a way out of the difficulties just described let us return back to the
expression $<n\mid\hat{O}\mid k>\exp(i\omega(n,k)t)$. Suppose that the algebra
of observables contains an identity element (operator). Then, by replacing
$\hat{O}$ \ by this operator \ we obtain, $<n\mid k>=<n(t)\mid k(t)>.$This
makes sense only if we require $<n\mid k>=const\delta_{nk}$. Clearly, we can
always adsorb the constant into the definition of the scalar product. In this
work, following [29], we suggest to replace the basic commutators, Eq.(20), by
the requirement of orthogonality. This requirement is compatible with the
requirement that the operators describing observables are Hermitian whose
eigenfunctions are mutually orthogonal. Instead of operators whose explicit
form is difficult to obtain we shall focus our attention on the properties of
orthogonal functions and, more generally, on the properties of orthogonal
polynomials (e.g. s$_{\lambda},$ etc.). Development of theory of orthogonal
polynomials of several variables in connection with quantum exactly solvable
model systems \ is an active area of current research [38,46,47]. Such an
approach makes sense since it is known [48] that \textsl{all} one- variable
orthogonal functions of exactly solvable problems in quantum mechanics [49]
are obtainable as various limiting cases of the Gauss-type hypergeometric
functions\footnote{This fact will be discussed in detail in Section 3.}.
Following ideas by Aomoto, Orlik and Terrao demonstrate that the
hypergeometric functions of multiple arguments (of which the Gauss-type is
just a special case) are expressible in the form of period
integrals\footnote{Periods can be associated with the homology basis
-different for different (algebraic) manifolds. Interested readers may consult
either [48] or [50] for more details.}. By the principle of complementarity
all many-body exactly solvable quantum mechanical problems should have
hypergeometric functions of multiple arguments as eigenfunctions. The most
important fact for our developments lies in the observation that when these
functions become eigenfunctions (as it is known in one component case), this
produces orthogonal polynomials-different for different many- body quantum
mechanical problems. This fact can be formulated as a problem : for a given
set of orthogonal polynomials find the corresponding many-body operator for
which such a set of orthogonal polynomials forms a complete set of eigenfunctions.

After these general remarks, we are ready to provide \ more concrete evidence
that this is indeed the case. The symmetric group $S_{n}$ has the following
presentation in terms of generators $s_{i}$ and Coxeter relations:%
\begin{align}
s_{i}^{2}  &  =1,\nonumber\\
s_{i}s_{j}  &  =s_{j}s_{i}\text{ \ for \ }\left\vert i-j\right\vert
\geq2,\nonumber\\
s_{i}s_{i+1}s_{i}  &  =s_{i+1}s_{i}s_{i+1}. \tag{21}%
\end{align}
If there is a set of $n$ elements of whatever kind the generator s$_{i}$
interchanges an element $i$ with $i+1$ so that \textit{s}$_{1},...,$%
\textit{s}$_{n-1}$ generate $S_{n}.$There are $n!$ permutations in the set of
$n$ elements. If we assign the initial ordered state, then any other state can
be reached by successful application of permutational generators to this
state. The word $w=$\textit{s}$_{a_{1}}$\textit{s}$_{a_{2}}\cdot\cdot\cdot
$\textit{s}$_{a_{l}}$ (where the indices $a_{1},...,a_{l}$ represent a subset
of the set of $n-1$ elements) can be identified with such a state. Since one
can reach this state in many ways, it makes sense to introduce the
\textsl{reduced word }\textit{w} whose length $\ l$($w$) is minimal. We would
like the generators of $S_{n}$ to act on monomials \textbf{x}$^{\mathbf{a}%
}=x_{1}^{a_{1}}x_{2}^{a_{2}}\cdot\cdot\cdot x_{n}^{a_{n}}$. For this purpose,
following Lascoux and Sch\"{u}tzenberger [51] (L-S) we introduce an operator
$\partial_{i}$ via rule:%
\begin{equation}
\partial_{i}:=\frac{1-s_{i}}{x_{i}-x_{i+1}}. \tag{22}%
\end{equation}
It acts on monomials such as \textbf{x}$^{\mathbf{a}}$ in such a way that the
generator $s_{i}$ acting on the combination $x_{i}^{a_{i}}x_{i+1}^{a_{i+1}}$
converts it into $x_{i}^{a_{i+1}}x_{i+1}^{a_{i}}.$ By construction, the action
of this operator on monomial is zero if $a_{i}=a_{i+1},$ otherwise it
diminishes the degree of the monomial by 1. In addition, the same authors
introduce operators%
\begin{equation}
\bar{\pi}_{i}=\frac{(1-s_{i})}{x_{i}-x_{i+1}}x_{i+1} \tag{23}%
\end{equation}
and $\pi_{i}=1+\bar{\pi}_{i}.$ Finally, \ being armed with such definitions,
we can introduce an operator $D_{i}(p,q,r)=p\partial_{i}+q\bar{\pi}_{i}%
+rs_{i}\footnote{By doing so, the operators $\partial_{i},\bar{\pi}_{i}$ and
$s_{i}$ become equivalent in the sense which we shall explain shortly.},$
where $p,q$ and $r $ are some numbers. L-S demonstrate that such defined
operator, while acting on monomials, obeys the braid-type relations (the 2nd
and the 3-rd lines in Eq.(21)) while the relation $s_i^2=1$ is replaced by
\begin{equation}
D_{i}^{2}=qD_{i}+r(q+r). \tag{24a}%
\end{equation}
With constants $p,q$ and $r$ properly chosen, such a relationship defines the
Hecke algebra $H_n$ of the symmetric group $S_n.$ Usually, it is written as%
\begin{equation}
D_{i}^{2}=(1-Q)D_{i}+Q \tag{24b}%
\end{equation}
with $Q$ being some number (effectively playing the same role as $p,q,r$).
$H_n$ should be considered as a deformation of $S_n$.The rationale for its
introduction lies in its direct connections with the knot and link theory so
that quantum mechanics can be considered as some branch of this theory
(Kholodenko 2006a). This fact will have its impact on quantization. To
demonstrate this, following Kirillov\ Jr.[52], by relabeling earlier defined
operator $\partial_i$ as $b_ij$ we reserve the notation $\partial
_i=\frac{\partial}{\partial x_i}$ for the usual operator of differentiation.
With its help we introduce the so called Dunkl operator $\mathcal{D}_i$ via%
\begin{equation}
\mathcal{D}_{i}=\partial_{i}+k%
{\textstyle\sum\limits_{j\neq i}}
b_{ij} \tag{25}%
\end{equation}
with $k$ being some (known) constant. Such defined operator acts on monomials
(polynomials). It possesses the property $w\mathcal{D}_iw^-1$=$\mathcal{D}%
_w(i)$ $\forall i\in S_n.$ Consider now the commutator [$\mathcal{D}%
_i$,$\mathcal{D}_j$]. Kirillov demonstrated that such a commutator is zero if
$b_ij$ satisfy the classical Yang-Baxter equation (CYBE)%
\begin{equation}
\lbrack b_{12},b_{13}]+[b_{12},b_{23}]+[b_{13},b_{23}]=0. \tag{26}%
\end{equation}
Alternatively, Eq.(26) can be taken as the definition for $b_ij$. This is
facilitated by designing of the degenerate affine Hecke algebra (Cherednik
2005).\ The purpose of this algebra from the physical point of view is to
introduce the Heisenberg commutation rule Eq.(20) without reference to the B-S
quanization prescription or to the (optical) sum rule described in Appendix A.
Such an algebra is made up as a semidirect product of $S_n$ with the
commutator algebra%
\begin{equation}
x_{i_{+1}}s_{i}-s_{i}x_{i}=h;\text{ }x_{i}s_{j}=s_{j}x_{i}\text{ \ }\forall
i\neq j,j+1;\text{ }x_{i}x_{j}=x_{j}x_{i}, \tag{27a}%
\end{equation}
where the constant $h$ is playing essentially the same role as the Plank's
constant $\hbar.$ From the \ above definitions it follows that Eq.(27a) is the
\ discrete analog of the Heisenberg's commutation rule, Eq.(20). Furthermore,
in view of the remark made after introduction of $D_i(p,q,r),$ it is possible
to rewrite the commutator in Eq.(27a) in the equivalent form. This indeed was
accomplished in the paper \ by Adin et al [53]. Hence, we can rewrite Eq.(27a)
equivalently as

\textsl{\bigskip}%
\begin{equation}
x_{i}\partial_{i}-\partial_{i}x_{i+1}=h;\text{ \ \ }\partial_{i}%
x_{i}\text{\ -\ \ }x_{i+1}\partial_{i}\text{\ }=h\text{\ ;\ \ }x_{i}%
\partial_{j}=\partial_{j}x_{i}\text{ \ }\forall i\neq j,j+1;\text{ }x_{i}%
x_{j}=x_{j}x_{i}, \tag{27b}%
\end{equation}
where $\partial_{i}$ should be understood in the sense of Eq.(22). At this
point it is useful to introduce yet another operator $\hat{s}_{i}%
=s_{i}+hb_{i,i+1}.$ It is designed in such a way that it obeys the braid
relations:%
\begin{equation}
\hat{s}_{1}\hat{s}_{2}\hat{s}_{1}=\hat{s}_{2}\hat{s}_{1}\hat{s}_{2}. \tag{28}%
\end{equation}
Furthermore, if now we define the operators $R_{12}=s_{1}\hat{s}_{1}%
,R_{23}=s_{2}\hat{s}_{2},R_{13}=s_{1}R_{23}s_{1}=s_{2}R_{12}s_{2},$ then the
Eq.(28) becomes equivalent to the standard Yang-Baxter (Y-B) equation for
$R_{ij}=1+hb_{ij}$ (or $R_{ij}\simeq\exp(hb_{ij})$ for $h\rightarrow0).$
Explicitly, we obtain: $R_{12}R_{13}R_{23}=R_{23}R_{13}R_{12}.$

All this discussion looks a bit formal at this point. Indeed, why to introduce
the operator $\mathcal{D}_{i}?$ Why to be concerned about the commutator
$[D_{i},D_{j}]?$ What the Yang-Baxter equations have to do with the results of
this paper? We would like to provide the answers to these questions now and in
the following subsection.

First, consider an equation $\mathcal{D}_{i}f=0.$ It can be written
alternatively as
\begin{equation}
\kappa\frac{\partial}{\partial z_{i}}f(\mathbf{z})=%
{\textstyle\sum\limits_{j\neq i}}
\frac{\Omega_{ij}}{z_{i}-z_{j}}f(\mathbf{z}) \tag{29}%
\end{equation}
which is just the celebrated Knizhnik-Zamolodchikov (K-Z) equation\footnote{In
fact, in general case [48] scalar function\textbf{\ }$f(\mathbf{z}%
)$\textbf{\ }is replaced by the vector function \textbf{\ }$\mathbf{f}%
(\mathbf{z}).$This fact should be kept in mind in actual calculations.}. This
means that: a) the operator $\mathcal{D}_{i}$ is effectively a covariant
derivative (the Gauss-Manin connection in the formalism of fiber bundles) and,
b) that the vanishing of commutator $[\mathcal{D}_{i},\mathcal{D}_{j}]$ is
just the zero curvature condition [54] essential for all known exactly
integrable systems. The question still remains: how $\Omega_{ij}$ in Eq.(29)
is related to $b_{ij}$ in Eq.(25)? The answer was found by Belavin and
Drinfel'd [55]. In the simplest (rational) case we have $b_{ij}(z)=\frac
{\Omega_{ij}}{z},$ as expected. More complicated trigonometric and elliptic
\ cases found by Belavin and Drinfel'd are summarized in the book by Etingof
with collaborators [56]. From the references just provided, it should be clear
that since solutions of the K-Z equations are expressible in terms of
hypergeometric functions of single and multiple arguments, all examples of
exactly solvable quantum mechanical problems (including those involving the
Dirac equation, and, hence, the spin) found in the textbooks on quantum
mechanics are covered by the formalism we have just described. In the next
subsection we would like to illustrate these results by concrete physical
examples taken from current physical literature.

\subsection{Latest developments in atomic physics illustrating general
principles}

\bigskip

In the review paper by Tanner et al [57] as well as in the book by
Cvitanovic$^{\prime}[58]$ it is explained in detail that in order to calculate
the He spectrum it is sufficient either: a) to consider the classical dynamics
of two electrons and the nucleus \textsl{on the line} and to use this
information in the semiclassical trace formula producing very accurate results
for the spectrum or, b) to restrict quantum mechanical calculations to the
spherical approximation (the so called s- wave approximation) in order to
arrive at the exactly solvable \textsl{radial} Schrodinger-type equation for
two electrons and massive nucleus[59,60] producing very reasonable results for
the He spectrum. To these achievements we would like to add those by
Svidzinsky et al [10,11] and Muravski and Swidzinsky [12] where the same type
or even better accuracy for He and other atoms and diatomic molecules is
obtained using the so called d-scaling. In this method the multielectron
Schr\"{o}dinger equation is analyzed in various dimensions. Upon proper
rescaling, the limiting case: d$\rightarrow\infty,$ is reduced again to the
exactly solvable radial-type multielectron equation which \ in the present
case becomes classical equation considered already by Bohr in 1913. Thus
again, the zeroth order exactly solvable problem is one dimensional.
Corrections in powers of 1/d are easily calculable producing results which
compare extremely well with much more cumbersome (and time consuming)
Hartree-Fock type calculations. To this list of examples it is appropriate to
add work by Ostrovsky and Prudov [61] which uses essentially the same
averaging and perturbation methods as developed in celestial mechanics [62]
superimposed with the Bohr-Sommerfeld quantization prescription. All examples
discussed in this subsection were done without theoretical guidance (other
than the proof of the existence of minimizers for Bohr-type functionals [13].
The theoretical framework developed in this section naturally explains \ why
these results are actually working so well. This framework sets up the stage
for developing applications of these results to celestial mechanics to be
discussed in the rest of this paper.

\section{Space, time and space-time in classical and quantum mechanics}

\subsection{General comments}

If one contemplates quantization of dynamics of celestial objects using
traditional textbook prescriptions, one will run into myriad of small and
large problems immediately. Unlike atomic systems in which all electrons repel
each other, have the same masses and \ are indistinguishable, in the case of,
say, Solar System all planets (and satellites) attract each other, have
different masses and visibly distinguishable. Besides, in the case of atomic
systems the Planck constant $\hbar$ plays prominent role while no such a role
can be given to the Planck constant in the sky. \ The only thing which remains
in common between both atomic and celestial dynamic systems is the existence
of stable closed orbits. In the previous section we demonstrated that this
fact is absolutely essential for quantization. Nevertheless, \ the formalism
developed thus far resembles more the existence theorem in mathematics rather
then the actual manual describing the computational protocol. The task now
lies in developing necessary constructive steps leading to actual
implementation of general principles. This task is accomplished below and in
the following section

\subsection{Space and time in classical and quantum mechanics}

Although \ celestial mechanics based on the Newton's law of gravity \ is
considered to be classical (i.e.nonquantum), \ with such an assumption one
easily runs into serious problem. Indeed, such an assumption \ implies that
the speed with which the interaction propagates is infinite and that the time
is the same everywhere. Wether this is true or false can be decided only
experimentally. Since at scales of our Solar System one has to use radio
signals to check correctness of Newton's celestial mechanics, one is faced
immediately with all kind of wave mechanics effects such as retardation, the
Doppler effect, etc. Because of this, the measurements are necessarily having
some error margins. The error margins naturally will be larger for more
distant objects. Accordingly, even at the level of classical mechanics applied
to the motion of celestial bodies we have to deal with certain inaccuracies
similar in nature to those in atomic mechanics. To make formalisms of both
atomic and celestial mechanics look the same one has to think about the space,
time and space-time transformations already at the level of classical mechanics.

We begin with observation that in the traditional precursor of quantum
mechanics-the Hamiltonian mechanics-the Hamiltonian equations \textsl{by
design} remain invariant with respect to the canonical transformations [63].
That is if \ sets $\{q_{i}\}$ and \{$p_{i}\}$ represent the "old" canonical
coordinates and momenta while $Q_{i}=Q_{i}(\{q_{i}\},\{p_{i}\})$ and
$P_{i}=P_{i}(\{q_{i}\},\{p_{i}\})$, $i=1-N$, represent the "new" set of
canonical coordinates and momenta, the Hamiltonian equations in the old
variables given by
\begin{equation}
\dot{q}_{i}=\frac{\partial H}{\partial p_{i}}\text{ \ and }\dot{p}_{i}%
=-\frac{\partial H}{\partial q_{i}} \tag{30}%
\end{equation}
and those rewritten in "new" \ variables will have the same form. Here we used
the commonly accepted notations, e.g. $\dot{q}_{i}=\frac{d}{dt}q_{i}$ , etc.
Quantum mechanics uses this form-invariance essentially since the Poisson
brackets introduced in Eq.(9) by design will \ also have the same form in
terms of both "old" and "new" canonical variables.

We would like to complicate matters by investigating the possibility of the
"canonical " time changes in classical mechanics. Fortunately, such a
possibility was explored to a some extent already. This is described in the
monograph by Pars [63] thus making our task considerably simpler. For the sake
of space, we refer our readers to pages 535-540 of this monograph.
Furthermore, following Dirac [64]$,$ we notice that a good quantization
procedure should always begin with the Lagrangian formulation of mechanics
since it is not always possible to make a transition from the Lagrangian to
Hamiltonian form of mechanics (and, thus, to quantum mechanics) due to the
presence of some essential constraints ( typical for mechanics of gauge
fields, etc.). Hence, we also begin with the Lagrangian functional
$\mathcal{L=L(}\{q_{i}\},\{\dot{q}_{i}\})$. The Lagrangian equations of motion
can be written in the form of Newton's equations given by $\dot{p}_{i}=F_{i},$
where the generalized momenta $p_{i}$ are given by $p_{i}=\delta
\mathcal{L}/\delta\dot{q}_{i}$ and the generalized forces $F_{i}$ are given by
$F_{i}=-\delta\mathcal{L}/\delta q_{i}.$ In the case if the total energy $E$
is conserved, it is possible instead of "real" time $t$ to introduce the
fictitious time $\theta$ via relation $dt=u(\{q_{i}\})d\theta$ where the
function $u(\{q_{i}\})$ is assumed to be nonnegative and is sufficiently
differentiable with respect to its arguments. At this point we can enquire if
Newton's equations can be written in terms of new time variable so that they
remain form- invariant. To do so, following Pars, we must: \ a) to replace
$\mathcal{L}$ by $u\mathcal{L},$ b) \ to replace $\dot{q}_{i}$ by
$\ q_{i}^{\prime}$ $/u,$ where $q_{i}^{\prime}$=$\frac{d}{d\theta}q_{i}$, c)
to rewrite the new Lagrangian in terms of \ such defined new time variables
and, finally, d) to obtain Newton's equations according to the described
rules, provided that now we have to use $p_{i}^{\prime}$ instead of $\dot
{p}_{i}$. In the case if the total energy of the system is conserved, we shall
obtain back the same form of Newton's equations rewritten in terms of new
variables. This means that by going from the Lagrangian to Hamiltonian
formalism of classical mechanics we can write the Hamilton's Eq.(30) in which
the dotted variables are replaced by primed. These arguments demonstrate
connections between space and time already at the level of classical
mechanics. Situation here is similar to that encountered in thermodynamics
where instead of absolute temperature one can use any nonegative function of
absolute temperature as new temperature. Using these arguments we notice that
since the temperature is conjugate to energy in thermodynamics, the time is
conjugate to energy in mechanics and, accordingly, in quantum mechanics. This
means that for the nondissipative (i.e. energy conserving) Hamiltonian
system\footnote{It should be kept in mind that the concept of nondissipativity
is actually of quantum origin (e.g. recall superconductors or superfluids). In
classical mechanics such a concept is just a convenient idealization similar
to the notion of a \ material point in Newton's mechanics or the notion of
thermodynamics when it is applied to the real heat engines, etc. The truly
nondissipative mechanical systems thus should behave quantum mechanically.
This observation provides the hint that some stable motions in our Solar
System are of quantum nature. In view of Eq.(6) this option makes sense.} the
Hamiltonian equations of motion, Eq.(30), will remain form- invariant if we
replace the Hamiltonian $H$ by some nonnegative function $f(H)$ while changing
time $t$ to time $\theta$ according to the rule $d\theta/dt=df(H)/dH\mid
_{H=E}$. Such a change will affect the quantum mechanics where now the
Schr\"{o}dinger's equation%
\begin{equation}
i\hbar\frac{\partial}{\partial t}\Psi=\hat{H}\Psi\tag{31a}%
\end{equation}
is to be replaced by
\begin{equation}
i\hbar\frac{\partial}{\partial\theta}\Psi=f(\hat{H})\Psi. \tag{31b}%
\end{equation}
With such an information at our hands, we would like to discuss the extent to
which symmetries of our (empty) space-time affect dynamics of particles
"living" in it.

\subsection{Space-time in quantum mechanics}

\subsubsection{\bigskip General comments}

Use of group-theoretic methods in quantum mechanics had began almost
immediately after its birth. It was initiated by Pauli in1926. He obtained a
complete quantum mechanical solution for\ the Hydrogen atom \ employing
symmetry arguments. His efforts were not left without appreciation. Our
readers can find many historically important references in two comprehensive
review papers by Bander and Itzykson [65]. In this subsection we pose and
solve the following problem: Provided that the symmetry of (classical or
quantum) system is known, will this information be sufficient for
determination of this system uniquely? Below, we shall provide simple and
concrete examples illustrating meaning of the word "determination". In the
case of quantum mechanics this problem is known as the problem about hearing
of the "shape of the drum". It was formulated by Mark Kac [66]. The problem
can be formulated as follows. Suppose that the spectrum of the drum is known,
will such an information determine the shape of the drum uniquely? The answer
is "No" [67]. Our readers may argue at this point that nonuniqueness could
come as result of our incomplete knowledge of symmetry or, may be, as result
of the actual lack of true symmetry (e.g. the Jahn-Teller effect in molecules,
etc. in the \ case of quantum \ mechanics). \ These factors do play some role
but they cannot be considered as decisive as the basic example below demonstrates.

\subsubsection{ Difficulties with the correspondence principle for Hydrogen
atom}

In this subsection we even do not use arguments by Kac. \ Since our arguments
are straightforward, they are more intuitively appealing. We choose the most
studied case of Hydrogen atom as an example.

As it is well known, the classical mechanical problem about motion of the
particle in centrally symmetric field is planar \ and is exactly solvable for
both the scattering and bound states [63,68]\textbf{. }The result of such a
solution depends on two parameters: the energy and the\textbf{\ }%
angular\textbf{\ }momentum. The correspondence principle formulated by Bohr is
expected to provide the bridge between the classical and quantum realities by
requiring that in the limit of large quantum numbers the results of quantum
and classical calculations for observables should coincide. Appendix A
provides a \ good example of such kind of thinking. However, this requirement
may or may not be possible to implement. It is violated already for the
Hydrogen atom. Indeed, according to the naive canonical quantization
prescriptions, one should begin with the \textsl{classical} Hamiltonian in
which one has to replace the momenta and coordinates by their operator
analogs. Next, one uses such constructed "quantum" Hamiltonian in the
Schr\"{o}dinger's equation, etc. Such a procedure breaks down at once for the
Hamiltonian of Hydrogen atom since the intrinsic planarity of the classical
\ Kepler's problem is entirely ignored thus leaving the projection of the
angular momentum without its classical analog. Accordingly, the scattering
states of Hydrogen atom \ and the classical mechanically obtained Rutherford's
formula \ obtained for planar configurations are reproduced quantum
mechanically (within the 1st Born approximation) using the 3-d
Schr\"{o}dinger's equation ! Thus, even for the Hydrogen atom the classical
and the quantum (or, better, pre quantum) Hamiltonians\textbf{\ }\textsl{do
not} match thus formally violating the correspondence principle. Evidently,
semiclassically we can only think of energy and the angular momentum thus
leaving the angular momentum projection unobserved. Such a "sacrifice" is
justified by the agreement between the observed and predicted Hydrogen atom
spectra and by use of Hydrogen-like atomic orbitals for multielectron atoms.
Although, to our knowledge, such a mismatch is not mentioned in any of the
students textbooks on quantum mechanics, its existence is essential if we are
interested in applications of quantum mechanical ideas to Solar System
dynamics. In view of such an interest, we would like to reconsider traditional
treatments of Hydrogen atom, this time being guided \ only by symmetry
considerations. This is accomplished in the next subsection.

\subsubsection{Emergence of the SO(2,1) symmetry group}

In April of 1940 Jauch and Hill\textbf{\ [}69\textbf{]} published a paper in
which they studied the\ planar Kepler problem quantum mechanically. Their work
was stimulated by earlier works by Fock of 1935 and by Bargmann of 1936 in
which it was shown that the spectrum of bound states for the Hydrogen atom can
be obtained by using representation theory of SO(4) group of rigid rotations
of 4-dimensional Euclidean space while the spectrum of scattering states can
be obtained by using the Lorentzian group SO(3,1). By adopting results of Fock
and Bargmann to the planar configuration Jauch and Hill obtained the
anticipated result. In the planar case one should use SO(3) group for the
bound states and SO(2,1) group for the scattering states. Although \ this
result will be reconsidered almost entirely, we mention about it now having
several purposes in mind.

First, we would like to reverse arguments leading to the final results of
Jauch and Hill in order to return to the problem posed at the beginning of
this section. That is, the fact that the Kepler problem is planar (due to
central symmetry of the force field) and the fact that the motion is
restricted to the plane and takes place in (locally) Lorentzian space-time are
the most general symmetry constraints imaginable. Thus, the fact that the
Lorentz SO(2,1) group is related to the spectrum of Kepler problem should be
anticipated. Nevertheless, the question remains: is Kepler's problem the only
one exactly solvable classical and quantum mechanical problem associated with
the SO(2,1) group? Below we demonstrate that, unfortunately, this is not the
case. In \ anticipation of such negative result, we would like to develop our
intuition by using some known results from quantum mechanics.

\subsubsection{\bigskip Classical-quantum correspondence allowed by SO(2,1)
symmetry: a gentle introduction}

For the sake of space, we consider here only the most generic (for this work)
example in some detail: the radial Schr\"{o}dinger equation for the planar
Kepler problem with the Coulombic potential. It is given by\footnote{The
rationale for discussing the Coulombic potential instead of gravitational will
be fully explained in the next section.}%
\begin{equation}
-\frac{\hbar^{2}}{2\mu}(\frac{d^{2}}{d\rho^{2}}+\frac{1}{\rho}\frac{d}{d\rho
}-\frac{m^{2}}{\rho^{2}})\Psi(\rho)-\frac{Ze^{2}}{\rho}=E\Psi(\rho). \tag{32}%
\end{equation}
Here $\left\vert m\right\vert =0,1,2,..$ is the angular momentum quantum
number as required. For $E<0$ it is convenient to introduce the dimensionless
variable $x$ via $\rho=ax$ and to introduce the new wave function: $\psi
(\rho)=\sqrt{\rho}\Psi(\rho)$. Next, by the appropriate choice of constant $a$
and by redefining $\psi(\rho)$ as $\psi(\rho)=\gamma x^{\frac{1}{2}+\left\vert
m\right\vert }\exp(-y)\varphi(y),$ where $y=\gamma x,$ -$\gamma^{2}=\frac{2\mu
E}{\hbar^{2}}a^{2},a=\frac{\hbar^{2}}{\mu ZE},$ the following hypergeometric
equation can be eventually obtained:%
\begin{equation}
\left\{  y\frac{d^{2}}{dy^{2}}+2[\left\vert m\right\vert +\frac{1}{2}%
-y]\frac{d}{dy}+2[\frac{1}{\gamma}-\left\vert m\right\vert -\frac{1}%
{2}]\right\}  \varphi(y)=0. \tag{33}%
\end{equation}
Formal solution of such an equation can be written \ as $\varphi
(y)=\mathcal{F}(-A(m),B(m),y),$ where $\mathcal{F}$ is the confluent
hypergeometric function. Physical requirements imposed on this function reduce
it to a polynomial leading to the spectrum of planar Kepler problem.
Furthermore, by looking into standard textbooks on quantum mechanics, one can
easily find that \textsl{exactly the same type of hypergeometric equation} is
obtained for problems such as one-dimensional Schr\"{o}dinger's equation with
the Morse-type potential,\footnote{That is, $V(x)=A(exp(-2\alpha
x)-2exp(-\alpha x)).$} three dimensional radial Schr\"{o}dinger equation for
the harmonic oscillator\footnote{That is, $V(r)=\dfrac{A}{r^{2}}+Br^{2}.$} and
even three dimensional radial equation for the Hydrogen atom\footnote{That is,
$V(r)=\dfrac{A}{r^{2}}-\dfrac{B}{r}.$}. Since the two-dimensional Kepler
problem is solvable with help of the representations of SO(2,1) \ Lorentz
group, the same should be true for all quantum problems just listed. That this
is the case is demonstrated, for example, in the book by Wybourne [70]. A
sketch of the proof is provided in Appendix B. This proof indicates that,
actually, the \textsl{discrete spectrum} of all problems just listed is
obtainable with help of SO(2,1) group. The question remains: if the method
outlined in Appendix B provides the spectra of several quantum mechanical
problems listed above, can we be sure that these are the only exactly solvable
quantum mechanical problems associated with the SO(2,1) Lorentz group?
\ Unfortunately, the answer is "No". More details are given below.

\subsubsection{\bigskip Common properties of quantum mechanical problems
related to SO(2,1) Lorentz group}

\bigskip

In Appendix B we provide a sketch of the so called spectrum-generating
algebras (SGA) method producing the exactly solvable one-variable quantum
mechanical problems. In this subsection we would like to put these results in
a broader perspective. In particular, in Section 2 we demonstrated
that\textsl{\ all exactly} \textsl{solvable quantum mechanical problem should
involve hypergeometric functions of single or multiple arguments}. We argued
that the difference between different problems can be understood topologically
in view of the discussed relationship with braid groups. On another hand,
obtained results, even though rigorous, are not well adapted for immediate
practical use. In this regard more useful would be to solve the following
problem: \textsl{For a given} \textsl{set of orthogonal polynomials find the
corresponding many-body operator for which such a set of orthogonal
polynomials forms a complete set of} \textsl{eigenfunctions}. At the level of
orthogonal polynomials of one variable relevant for all exactly solvable
two-body problems of quantum mechanics, one can think about related problem of
finding all potentials in the one-dimensional radial Schr\"{o}dinger equation,
e.g. Eq.(B.1), leading to the hypergeometric-type solutions. Such a task was
accomplished by Natanzon [71]. Subsequently, his results were reinvestigated
by many authors with help of different methods, including SGA. To our
knowledge, the most complete recent summary of the results, including
potentials and spectra can be found in the paper by Levai [72]. Even this
(very comprehensive) paper does not cover all aspects of the problem. For
instance, it does not mention the fact that these results had been extended to
relativistic equations such as Dirac and Klein-Gordon for which similar
analysis was made by Cordero with collaborators [73] . In all cited cases
(relativistic and non relativistic) the underlying symmetry group was SO(2,1).
The results of Appendix B as well as all other listed references can be traced
back to the classically written papers by Bargmann [74] and Barut and Fronsdal
[75] on representations of SO(2,1) Lorentz group. Furthermore, the discovered
connection \ of this problematic with supersymmetric quantum mechanics [76,77]
can be traced back to the 19th century works by Gaston Darboux \ [72].

Summarizing, established in Section 2 rigorous connections between exactly
solvable two-body quantum mechanical problems and hypergeometric functions
and, by complementarity principle, between exactly solvable many body problems
and hypergeometric functions of many arguments are consequences of the locally
Lorentzian group structure of our space-time. Such a structure allows many
\textsl{but not infinitely many} exactly solvable problems to exist. The fact
that \textsl{planar} SO(2,1) is sufficient to cover all known exactly solvable
two-body cases (instead of the full SO(3,1) Lorentz group!) is quite
remarkable. It is sufficient for the purposes of this work but leaves open the
question : Will use of the full Lorentz group lead to the exactly solvable
quantum mechanical problems not accounted by SO(2,1) group symmetry? \ This
topic \ will be discussed in Section 5. In the meantime, we would like to
address the problem of quantization of Solar System dynamics using results of
Sections 2 and 3. This is done in the next section

\section{\ Quantum celestial mechanics of Solar System}

\subsection{ General remarks}

We begin this section by returning back to Eq.(6) once again. Based on
previous discussions, this equation provides us with the opportunity to think
seriously about quantum nature of our Solar System dynamics. Nevertheless,
such an equation reveals only one aspect of quantization problem and, as such,
provides only a sufficient condition for quantization. The necessary condition
in atomic and celestial mechanics lies in the \textsl{nondissipativity} of the
dynamical systems in both cases\footnote{E.g. see the paper by Goldreich [25]
mentioned in Section 1.2.}. Recall that Bohr introduced his quantization
prescription to avoid dissipation caused by the emission of radiation \ by
electrons in orbits in general position. As we demonstrated previously, new
quantum mechanics have \textsl{not} explained absence of dissipation for
stationary Bohr's orbits\footnote{At the level of old Bohr theory absence of
dissipation at the stationary Bohr orbit was explained by Boyer [78].
Subsequently his result was refined by Puthoff\textbf{\ [}79\textbf{]}.}. In
fact, as our analysis of Heisenberg's work(s) indicates, new quantum mechanics
have \textsl{not} added a single new element to the old atomic mechanics in
terms of the new issues to be considered.

In the nutshell, \textsl{new quantum mechanics provided a} \textsl{convenient
computational scheme for dealing with otherwise purely mechanical problems
involving accidental degeneracy (that is resonances)}. \textsl{By}
\textsl{doing so, it made no attempt at explaining (using \ known results from
mechanics and electrodynamics) the nondissipativity. }Nevertheless, the
phenomenon of nondissipativity was explained quite convincingly in the case of
superconductivity and superfluidity later on. Thanks to these intrinsically
quantum phenomena, we can be sure that quantum mechanics did capture some
truth. Regrettably, only some since, as we discussed in Section 3.2.2, even
for the most studied case of Hydrogen atom the task of establishing the
correspondence between the classical and quantum models of Hydrogen atom is
nontrivial. The symmetry (and supersymmetry) arguments of Section 3 based on
locally Lorentzian space-time structure as well as the combinatorial arguments
of Section 2 simplified task of establishing the quantum-classical
correspondence considerably. \ This happened because of firmly established
finite number of exactly solvable quantum mechanical problems allowed by the
Lorentzian-type symmetry whose spectra are known and documented. These facts
allow us to think\ seriously about quantization of Solar System dynamics.

\subsection{From Laplace to \ Poincare$^{\prime}$ and Einstein}

Before discussing this issue in some detail, we still need to make several
remarks. First, although superficially classical Hamiltonians for Coulombic
and Newtonian potentials look almost the same, the naive textbook-style
quantization will immediately run into major problems. For one thing, all
electron masses are the same while all planetary/satellite masses are
different. For other thing, filling of atoms by electrons is controlled by the
electric charge of the nucleus so that stable atoms/molecules are electrically
neutral. Apparently, no such restriction exists for the system of gravitating
bodies. Next, apparent violation of planarity of Hydrogen atom treated at the
level of classical mechanics is justified by the fact that the angular
momentum projection does play an important role in chemistry. As far as we can
see, nothing of that sort exists in the sky.

To deal with the mass differences \ for planetary systems we have to recall
some facts from general relativity. We shall restrict ourself only by some
illustrative examples meant to provide some feeling of problems we would like
to discuss. To this purpose we would like to make some comments on the
classical mechanical treatment of Kepler problem in representative physics
textbooks, e.g. read [68,80].Such treatments tend to ignore the equivalence
principle essential for the gravitational Kepler problem and nonexistent for
the Coulomb-type problems. This causes some significant inaccuracies to
emerge. Specifically, according to Vol.2 of famous Landau-Lifshitz course in
theoretical physics [81] if we take $\mathcal{L}$=$\dfrac{m\mathbf{v}^{2}}%
{2}-m\mathbf{\varphi}$ as the Lagrangian for a particle in gravitational field
(represented by a local potential $\mathbf{\varphi),}$the Lagrangian
(Newtonian) equations of motion can be written as
\begin{equation}
\mathbf{\dot{v}=-\nabla\varphi} \tag{34}%
\end{equation}
so that the mass drops out of this equation making it possible to think about
such an equation as an equation for geodesic in pseudo-Riemannian space. This
observation had lead Einstein to full development of general relativity
theory. By noticing that Newton's equation makes sense only for material
points ( that is for idealized \textsl{formally nonexisting} objects), the
same must be true for Eq.(34). Hence, as such it is valid only for the well
localized point-like objects. Using such idealized model, we need to discuss
briefly the 2-body Kepler problem for particles with masses $m_{1}$ and
$m_{2}$ interacting gravitationally. The Lagrangian for this problem is given
by%
\begin{equation}
\mathcal{L}=\frac{m_{1}}{2}\mathbf{\dot{r}}_{1}^{2}+\frac{m_{2}}%
{2}\mathbf{\dot{r}}_{2}^{2}+\gamma\frac{m_{1}m_{2}}{\left\vert \mathbf{r}%
_{1}-\mathbf{r}_{2}\right\vert }. \tag{35a}%
\end{equation}
Introducing the center of mass \ and relative coordinates via $m_{1}%
\mathbf{r}_{1}+m_{2}\mathbf{r}_{2}=0$ and $\mathbf{r}=\mathbf{r}%
_{1}-\mathbf{r}_{2},$ the above Lagrangian \ can be rewritten as%
\begin{equation}
\mathcal{L=}\frac{\mu}{2}\mathbf{\dot{r}}^{2}+\gamma\frac{m_{1}m_{2}%
}{\left\vert \mathbf{r}\right\vert }\equiv\frac{m_{1}m_{2}}{m_{1}+m_{2}}%
(\frac{\mathbf{\dot{r}}^{2}}{2}+\gamma\frac{(m_{1}+m_{2})}{\left\vert
\mathbf{r}\right\vert }), \tag{35b}%
\end{equation}
where, as usual, we set $\mu=\frac{m_{1}m_{2}}{m_{1}+m_{2}}.$The constant
$\frac{m_{1}m_{2}}{m_{1}+m_{2}}$ can be dropped and, after that, instead of
the geodesic (34) we obtain the equation for a fictitious point-like object of
unit mass moving in the field of gravity produced by the point-like body of
mass $m_{1}+m_{2}$. Clearly, in general, one cannot talk about geodesics in
this case. Nevertheless, as it is usually done, if, say, $m_{1}\gg m_{2\text{
}}$ (as for the electron in Hydrogen atom or for the Mercury rotating around
Sun) \ one can with very good accuracy discard mass $m_{2\text{ }}$ thus
obtaining an equation for a geodesic. Such an approximation was indeed made by
Einstein in his major work on general relativity [82] in which he ignored the
mass of Mercury entirely when making his calculations of the perihelium shift
for this planet. More recent results [83] show that such an approximation is
expected to be quite satisfactory for other planets of our Solar
System\footnote{E.g. read \ Box 40.3 of this reference as well as pages
1126-1129.}. With the exception of Pluto-Charon system, where $\mu_{2}$
$=m_{2}/(m_{1}+m_{2})$ is of order $10^{-1},$ and the Earth-Moon system, where
$\mu_{2}$ is of order $10^{-2}$, \ all other planet-satellite and Sun- planet
pairs have $\mu_{2}$ of order 10$^{-3\text{ \ }}$and less [26] so that use of
geodesics is justifiable physically. Mathematically, however, this is not
quite the case yet since, even in the case of Mercury considered by Einstein,
it is necessary to prove that influence of the rest of planets of Solar System
on its motion can be ignored (as well as finite size of the Sun, etc.).

The task of proving that motion of planets can be well approximated by
geodesics can be traced back to works by Laplace on celestial mechanics.
Lecture notes by Moser and Zehnder [84] contain accessible discussion of
Laplace's works\footnote{E.g. read pages 102-120.} to which we refer our
readers for details. In short, Laplace was interested in dynamics of the
planar 4-body problem using \ Jupiter and its 3 satellites: Io, Europe and
Ganymede, as an example. He noticed that the motion of these satellites obeys
the resonance condition and he was able to reproduce this motion analytically
by ignoring satellite masses (just like in Eq.(35b), but beginning with the
full 4-body problem initially !). Under these conditions, gravitational
interactions between satellites can be neglected so that the motions become
completely decupled but subject to the resonance condition. Furthermore, to
study stability of such resonance motions Laplace (and Lagrange) assumed that
the actual (Lagrangian) motions\footnote{E.g. see Arnol'd et al [18], page
261.} of satellites oscillate about the respective \ stable orbits of these
satellites. Thus, \textsl{effectively, Laplace and Lagrange were considering
the effects of general relativity} \textsl{and quantum mechanics long before
these disciplines have been officially inaugurated}. In their lectures, Moser
and Zehnder also provide references to works by Poincare$^{\prime}$ and de
Sitter on further refinements of Laplace's results. \ Although according to
Arnol'd et al [18] the extension of work by Laplace to the full n+1 body
planar problem was given in the monograph by Charlier [85], rigorous
mathematical proofs have been obtained only quite recently by Fejoz [86] and
Biasco et al [87]. To realize the difficulties in providing such a proof it is
sufficient, following Poincare$^{\prime}[88],$to demonstrate that the results
of massless limit considered by Laplace will remain practically unchanged if
the satellites would have some small but finite masses (so that they interact
with each other !). Such a philosophy lies at the heart of KAM theory used and
improved in the works by Fejoz [86] and Biasco et al [87].

Even with these proofs available,\ one should \ take into account that, in
view of experimental \ limitations, Newton's \ law of gravity should amended
rigorously speaking. This is so taking into account the finite speed of
propagation of gravitational interaction as well as the fact that all
observations are made with some kind of light/radio sources causing
retardation, Doppler and other effects. Thus, \ taking into account
experimental conditions, the traditional \textsl{classical} \textsl{mechanics}
\ description of celestial motions \ becomes replaced by that
\textsl{\ encountered in quantum mechanics} where one has to use probabilities
to account for incompleteness of available information. Furthermore, the above
proofs do not account for dissipation effects playing major stabilizing role
in both atomic and \ quantum celestial mechanics.

If \ we assume that the motion of bodies indeed takes place on geodesics then,
formally, there are no interactions and the local time becomes proper time. In
the case of, say, binary stars of comparable masses one cannot use geodesics
for description of their relative motion\footnote{This case was discussed in
papers by Einstein, Infeld and Hoffmann [89] and Robertson [90]\ with the
outcome that it is possible to describe gravitational field \textsl{outside}
such a binary system in terms of geodesics. This leaves open the question of
dynamical stability of such binaries since their motion is controlled by the
Newton's equations of motion. In view of the effects of tidal friction, which
should be quite appreciable in this case, the dynamics of such binaries should
be most likely unstable. For such systems one can safely neglect friction
caused by the emission of gravitational waves since these are effects of fifth
order in $c^{-1}$(e.g. read Landau and Lifshitz [81], paragraph 106).} so that
one is confronted with the problem of matching the Einsteinian \ gravity with
\ its Newtonian limit as discussed in the paper by Einstein, Infeld and
Hoffmann [89] \footnote{See also Section 6 below.}. In all other theories of
gravity, including the Brans-Dicke-Jordan's theory, there are substantial
departures from the geodesic motion. Details can be found on pages 1127-1129
of the book by Misner et al [83].

Clearly, the difficulties of explaining motions using classical mechanics of
n+1 body problem are such that the assumption about truly geodesic motion
\ looks suspicious. But these results are based on Newtonian mechanics which
by design do not account for dissipation and retardation effects. The facts we
just mentioned also complicate the choices between different (alternative)
theories of gravity. Hence, it is clear that at the present state of our
knowledge the ultimate choice between competing theories can be made only
based on additional information. Such an information is supplied, in part, in
this work where uses of historical analogies between the quantum (atomic)
\ and celestial mechanics provide some helpful guidance. For this purpose we
compiled our Table 3 prior to actual computations.

\mathstrut

\ \ \ \ \ \ \ \ \ \ \ \ \ \ \ \ \ \ \ \ \ \ \ \ \ \ \ \ \ \ \ \ \ \ \ \ \ \ \ \ \ \ \ \ \ \
\begin{tabular}
[c]{l}%
Table 3
\end{tabular}

\
\begin{tabular}
[c]{|c|c|c|}\hline%
\begin{tabular}
[c]{l}%
$\backslash$%
$Type$\textit{\ }$of$\textit{\ }$mechanics$\\
$\mathit{Properties}$%
\end{tabular}
&
\begin{tabular}
[c]{l}%
$Quantum$\textit{\ }$atomic$\\
$mechanics$%
\end{tabular}
&
\begin{tabular}
[c]{l}%
$\mathit{Quantum}$\\
$celestial$\textit{\ }$mechanics$%
\end{tabular}
\\\hline%
\begin{tabular}
[c]{l}%
Dissipation (type of)%
$\backslash$%
\\
(yes%
$\backslash$%
no)%
$\backslash$%
on stable orbits
\end{tabular}
&
\begin{tabular}
[c]{l}%
electromagnetic\\
friction%
$\backslash$%
no%
$\backslash$%
\\
Bohr orbits
\end{tabular}
&
\begin{tabular}
[c]{l}%
tidal friction\\%
$\backslash$%
no%
$\backslash$%
Einstein's geodesics
\end{tabular}
\\\hline%
\begin{tabular}
[c]{l}%
Accidental degeneracy%
$\backslash$%
\\
(yes%
$\backslash$%
no)%
$\backslash$%
origin
\end{tabular}
& yes%
$\backslash$%
Bohr-Sommerfeld condition & yes%
$\backslash$%
closure of the Lagrangian orbits\\\hline
Charge neutrality & yes & no(but see below)\\\hline
Masses &
\begin{tabular}
[c]{l}%
electrons having\\
the same masses
\end{tabular}
&
\begin{tabular}
[c]{l}%
(up to validity of the\\
equivalence principle)\\
masses are the same
\end{tabular}
\\\hline
Minimal symmetry group & SO(2,1) & SO(2,1)\\\hline%
\begin{tabular}
[c]{l}%
Correspondence\\
principle
\end{tabular}
& occasionally violated & occasionally violated\\\hline%
\begin{tabular}
[c]{l}%
Discrete spectrum:\\
finite or infinite%
$\backslash$%
reason%
$\backslash$%
\\
Pauli principle(yes%
$\backslash$%
no)
\end{tabular}
&
\begin{tabular}
[c]{l}%
finite and infinite%
$\backslash$%
\\
charge neutrality%
$\backslash$%
\\
yes
\end{tabular}
&
\begin{tabular}
[c]{l}%
finite%
$\backslash$%
\\
no charge neutrality%
$\backslash$%
\\
yes
\end{tabular}
\\\hline
\end{tabular}
\ \ \ \ \ 

Details related to this table are discussed further below.\ \ \ \ \ \ \ \ \ \ \ \ \ \ \ \ \ \ \ \ \ \ \ \ \ \ \ \ \ \ \ \ 

\subsection{ Celestial spectroscopy, the Titius-Bode law\ of planetary
distances and quantum celestial mechanics}

The atomic spectroscopy was inaugurated by Newton, in the second half of 17th
century. As we discussed in the Introduction, the results of atomic and
molecular spectroscopy were used by Bohr in essential way resulting in the
birth of quantum mechanics. The celestial spectroscopy was inaugurated by
Titius in the second half of 18th century and become famous after it was
advertised by Johann Bode, the Editor of the "Berlin Astronomical Year-book"
in the late 18th century. The book by Nieto [91] provides extensive
bibliography related to uses and interpretations of the Titius-Bode (T-B) law
up to second half of 20th century. Unlike the atomic spectroscopy, where the
observed atomic and molecular spectra were expressed using simple empirical
formulas which were (to our knowledge) never elevated to the status of
\ "law", in celestial mechanics the empirical T-B formula
\begin{equation}
r_{n}=0.4+03.\cdot2^{n}\text{, \ \ \ }n=-\infty,0,1,2,3,... \tag{36}%
\end{equation}
for the orbital radii (semimajor axes) of planets acquired the status of a law
in the following sense. In the case of atomic spectroscopy the empirical
formulas used for description of the atomic/molecular spectra have not been
used (to our knowledge) for making predictions. Their purpose was just to
describe in mathematical terms what had been observed. Since the T-B empirical
formula for planetary distances was used as the law, it was used in search for
planets not yet discovered. In such a way Ceres, Uranus, Neptune and Pluto
were found [92]. However, the discrepancies for Neptune and Pluto were much
larger than the error margins allowed by the T-B law\footnote{Chapter 10 of
[92] provides a very lively account of the present knowledge about various
objects "living" in the Solar System.}.This fact divided the astronomical
community into "believers" and "atheists" regarding to the meaning and uses of
this law. Without going into historical details, we would like to jump to the
very end of the Titius-Bode story in order to use its latest version \ which
we found in the paper by Neslu\v{s}an [93] who, in turn, was motivated by the
work of Lynch [94]. Instead of Eq.(36) these authors use another empirical
power law dependence
\begin{equation}
r_{n}=r_{0}B^{n},\text{ }n=1,2,3,..,9. \tag{37}%
\end{equation}
For planets (except Pluto and including the asteroid belt) Neslu\v{s}an
obtained\footnote{In astronomical units (to be defined below).} $r_{0}%
(au)=0.203$ and $B=1.773$ with the rms deviation accuracy of
0.0534\footnote{This result gives for the Earth in astronomical (au) units the
result $r_{3}\simeq1.13.$ Much better result is obtained in case if we choose
$B=1.7.$ In this case we obtain: $r_{3}\simeq.997339.$ Lynch (2003) provides
$B=1.706$ and $r_{0}=0.2139.$}. Analogous power law dependencies were obtained
previously in the work by Dermott [95] for both planets and satellites of
heavy planets such as Jupiter, Saturn and Uranus.

It should be noted that \ because of noticed discrepancies the attempts were
made to prove or disprove the Titius-Bode law by using statistical analysis,
e.g. see papers by Lynch [93] and Hayes and Tremaine [96], with purpose of
finding out to which extent the observed dependencies can be considered as non
accidental. \textsl{In view of Heisenberg quantization (honeycomb) condition,
Eq.(5.a), it should be obvious by now that whatever distribution of
frequencies can be} \textsl{measured, it can, in principle, lead to
quantization}. \textsl{In principle, because to implement this in practice
requires to identify possible models and the Hamiltonians for these models as
\ we discussed extensively in the previous sections}. Hence, in the present
case we are confronted \ with exactly the same task. To move forward \ some
historical analogies are helpful at this time.

When Bohr was analyzing the data for He atom (Table 1) he had in mind a model
of He made of two independent electrons rotating around the same nucleus. As
results of \ Section 1 indicate, such an approximation produced quite
reasonable results. Clearly, when dealing with dynamics of Solar System, one
would like to follow the same philosophy. That is to assume first that the
planets are noninteracting and move along the geodesics independently. In the
case of atomic mechanics it was clear from the beginning that such an
approximation should sooner or later fail even though it works well in some
cases. For exactly the same reasons it is rather naive to expect that the T-B
law makes always sense. Rather, it makes sense to assume that it works for as
long as the assumption of noninteracting planets moving on geodesics can be
checked quantum mechanically. Furthermore, the nonexisting electroneutrality
in the sky \ provides strong hint that the T-B law must be of very limited use
since the number of discrete levels \ for gravitating systems should be always
finite. Otherwise we would observe the countable infinity of satellites around
\ Sun or heavy planets which is both observationally and physically wrong. In
the literature one can find many attempts at quantization of Solar System
using sttandard prescriptions of quantum mechanics\footnote{Since this work is
not a review, we do not provide references to papers whose results do not
affect ours.}. Because of this, restrictions on the number of allowed discrete
levels cannot be made.

In the present case, to facilitate matters, we would like to make several
additional observations. First, we have to find the analog of the Planck
constant. Second, we have to have some mechanical model in mind to make our
search for correct answer meaningful. To accomplish the first task, we have to
take into account the 3-rd Kepler's law. In accord with Eq.(35b), it can be
written as $r_{n}^{3}/T_{n}^{2}=\dfrac{4\pi^{2}}{\gamma(M+m)}$. In view of
arguments presented in previous subsection, we can safely approximate this
result by $4\pi^{2}/\gamma M$, where $M$ is the mass of Sun. For the
\ purposes of this work, it is convenient to restate this law as
$3lnr_{n}-2\ln T_{n}=\ln4\pi^{2}/\gamma M=const$ \ \ Below, we choose the
\textsl{astronomical system of units} in which $4\pi^{2}/\gamma M=1.$ By
definition, in this system of units we have for the Earth: $r_{3}=T_{3}=1$.

Consider now the Bohr result, Eq.(4), and take into account that
$E=\hbar\omega\equiv\dfrac{h}{2\pi}\dfrac{2\pi}{T}.$Therefore, Bohr's result
can be conveniently restated as $\omega(n,m)=\omega(n)-\omega(m).$Taking into
account Eq.s(4),(31b),(37) and the third Kepler's law we obtain:%
\begin{equation}
\omega(n,m)=\frac{1}{c\ln\tilde{A}}(nc\ln\tilde{A}-mc\ln\tilde{A}), \tag{38}%
\end{equation}
where the role of Planck's constant is played now by $c\ln\tilde{A}$ where
$\tilde{A}=B^{\frac{3}{2}}$ and $c$ is some constant which will be determined
selfconsistently below\footnote{Not to be confused with the speed of light
!}$.$

At first, one may think that what we obtained is just a simple harmonic
oscillator spectrum. After all, this should come as not too big a surprise
since in terms of the action -angle variables all exactly integrable systems
are reducible to the sets of harmonic oscillators. This result is also
compatible with the results of Appendix B. The harmonic oscillator option is
physically undesirable in the present case since for gravitating systems the
charge \ neutrality constraint cannot be imposed, e.g. see Table 3. Evidently,
allowing such a spectrum is equivalent to the correctness of the T-B law. But
it is well known that this law is not working well for larger numbers. In
fact, it would be extremely strange should it be working in this regime since
the total mass of all harmonically bound planets could potentially become infinite.

To make a progress, we have to use the 3rd Kepler's law once again, i.e. we
have to take into account that in chosen astronomical system of units
$3lnr_{n}=$ $2\ln T_{n}.$ In view of the arguments just presented, a quick
look at Eq.s B(13),(14) suggests that the underlying mechanical system is
likely to be associated with that for the Morse potential. The low lying
states of such a system cannot be distinguished from those for the harmonic
oscillator. However, this system does have only a finite number of energy
levels which makes sense physically. The task remains to connect this system
with the planar Kepler's problem. Although in view of results of Appendix B
such a connection does indeed exist, we would like to demonstrate it
explicitly at the level of classical mechanics.

Following Pars [63], the motion of a point of unit mass in the field of
gravity is described by the following equation%
\begin{equation}
\dot{r}^{2}=(2Er^{2}+2\gamma Mr-\alpha^{2})/r^{2}, \tag{39}%
\end{equation}
where $\alpha$ is the angular momentum integral (e.g. see Eq.(5.2.55) of Pars
book). We would like now to replace $r(t)$ by $r(\theta)$ in such a way that
$dt=u(r$($\theta)$)$d\theta$ . Let therefore $r(\theta)=r_{0}\exp(x(\theta)),$
-$\infty<x<\infty.$ Unless otherwise specified, we shall write $r_{0}=1$. In
such (astronomical) system of units) we obtain, $\dot{r}=x^{\prime}%
\dfrac{d\theta}{dt}\exp(x(\theta)).$ This result can be further simplified by
choosing $\dfrac{d\theta}{dt}=\exp(-x(\theta)).$ With this choice Eq.(39)
acquires the following form:%
\begin{equation}
(x^{\prime})^{2}=2E+2\gamma M\exp(-x)-\alpha^{2}\exp(-2x). \tag{40}%
\end{equation}
Consider points of equilibria for the potential \ $U(r)=-2\gamma
Mr^{-1}+\alpha^{2}r^{-2}.$ From here we obtain: $r^{\ast}=\dfrac{\alpha^{2}%
}{\gamma M}.$ According to Goldstein et al\ [80] such defined $r^{\ast}$
coincides with the major elliptic semiaxis. It can be also shown, e.g. Pars,
Eq.(5.4.14), that for the Kepler problem the following relation holds:
$\ E=-\dfrac{\gamma M}{2r^{\ast}}$. Accordingly, $r^{\ast}=-\dfrac{\gamma
M}{2E}, $ and, furthermore, using condition $\frac{dU}{dr}=0$ we obtain,
$\dfrac{\alpha^{2}}{\gamma M}=-\dfrac{\gamma M}{2E}$ or, $\alpha^{2}$
$=-\dfrac{\left(  \gamma M\right)  ^{2}}{2E}.$ Since in the chosen system of
units $r(\theta)=\exp(x(\theta)),$ we obtain, $\dfrac{\alpha^{2}}{\gamma
M}=\exp(x^{\ast}(\theta)).$ It is convenient to choose $x^{\ast}(\theta)=0.$
This requirement makes the point $x^{\ast}(\theta)=0$ as the origin and
implies that with respect to such chosen origin $\alpha^{2}=\gamma
M\footnote{In doing so some caution should be exercised since upon
quantization equation $r^{\ast}=\dfrac{\alpha^{2}}{\gamma M}$ becomes
$r_{n}^{\ast}=\dfrac{\alpha_{n}^{2}}{\gamma M}.$ Selecting the astronomical
scale $r_{3}^{\ast}=1$ as the unit of length implies then that we can write
the angular momentum $\alpha_{n}^{2}$ as $\varkappa$ $\dfrac{r_{n}^{\ast}%
}{r_{3}^{\ast}}$ and \ to define $\varkappa$ as $\alpha_{3}^{2}$ $\equiv
\alpha^{2}.$}.$ Using this fact Eq.(40) can then be conveniently rewritten as
\begin{equation}
\frac{1}{2}(x^{\prime})^{2}-\gamma M(\exp(-x)-\frac{1}{2}\exp(-2x))=E
\tag{41a}%
\end{equation}
or, equivalently, as
\begin{equation}
\frac{p^{2}}{2}+A(\exp(-2x)-2\exp(-x))=E, \tag{41b}%
\end{equation}
where $A=\dfrac{\gamma M}{2}.$ \ \ Since this result is exact classical analog
of the quantum Morse potential problem, transition to quantum mechanics can be
done straightforwardly at this stage. By doing so \ we have to replace the
Planck's constant $\hbar$ by $c\ln\tilde{A}$. After that, we can write the
answer for spectrum at once [97]:%
\begin{equation}
-\tilde{E}_{n}=\frac{\gamma M}{2}[1-\frac{c\ln\tilde{A}}{\sqrt{\gamma M}%
}(n+\frac{1}{2})]^{2}. \tag{42}%
\end{equation}
This result contains an unknown parameter $c$ which we would like to determine
now. To do so it is sufficient to expand the potential in Eq.(41b) and to keep
terms up to quadratic. Such a procedure produces the anticipated harmonic
oscillator result%
\begin{equation}
\frac{p^{2}}{2}+Ax^{2}=\tilde{E} \tag{43}%
\end{equation}
with the spectrum given by $\tilde{E}_n=(n+\frac{1}{2})c\sqrt{2A}\ln\tilde
{A}.$ In the astronomical system of units the spectrum reads: $\tilde
{E}_n=(n+\frac{1}{2})c2\pi\ln\tilde{A}$ . This result is in agreement with
Eq.(38). To proceed, we \ notice that in Eq.(38) the actual sign of the
Planck-type constant is undetermined. Specifically, in our case (up to a
constant) the energy $\tilde{E}_n$ is determined by ln$\left(  \frac{1}%
{T_n}\right)  $ $=-\ln$ $\tilde{A}$ so that it makes sense to write $%
-\tilde{E}_n\sim n\ln\tilde{A}.$ To relate the classical energy defined by the
Kepler-type equation $E=-\dfrac{\gamma M}{2r^\ast}$ to the energy we just have
obtained, we have to replace this Kepler-type equation by $-\tilde{E}%
_n\equiv-\ln\left\vert E\right\vert =-2\ln\sqrt{2}\pi+\ln r_n\text{ \ }$This
is done in view of the 3rd Kepler's law and the fact that the new coordinate
$x$ is related to the old coordinate $r$ via $r=e^x$. Using Eq.(37) (for
$n=1$) in the previous equation and comparing it with already obtained
spectrum of harmonic oscillator we obtain:
\begin{equation}
-2\ln\sqrt{2}\pi+\ln r_{0}B=-c2\pi\ln\tilde{A}, \tag{44}%
\end{equation}
where in arriving at this result we had subtracted the nonphysical ground
state energy. Thus, we obtain:%
\begin{equation}
c=\frac{1}{2\pi\ln\tilde{A}}\ln\frac{2\pi^{2}}{r_{0}B}. \tag{45}%
\end{equation}
Substitution of this result back into Eq.(42) produces%
\begin{align}
-\tilde{E}_{n}  &  =2\pi^{2}[1-\frac{(n+\frac{1}{2})}{4\pi^{2}}\ln\left(
\frac{2\pi^{2}}{r_{0}B}\right)  ]^{2}\simeq2\pi^{2}[1-\frac{1}{9.87}%
(n+\frac{1}{2})]^{2}\nonumber\\
&  \simeq2\pi^{2}-4(n+\frac{1}{2})+0.2(n+\frac{1}{2})^{2}. \tag{46}%
\end{align}
To determine the number of bound states, we follow the same procedure as
developed in chemistry for the Morse potential. \ For this
purpose\footnote{Recall that in chemistry the Morse potential is being
routinely used for description of the vibrational spectra of diatomic
molecules.} we introduce the energy difference $\Delta\tilde{E}_{n}=$
$\tilde{E}_{n+1}-\tilde{E}_{n}=4-0.4(n+1)$ first. Next, the maximum number of
bound states is determined by requiring $\Delta\tilde{E}_{n}=0.$ In our case,
we obtain: $n_{\max}=9$. This number is in perfect accord with observable data
for planets of our Solar System (with Pluto being excluded and the asteroid
belt included). In spite of such a good accord, some caution must be exercised
while analyzing the obtained result. Should we not insist on physical grounds
that the discrete spectrum must contain only finite number of levels, the
obtained spectrum for the harmonic oscillator would be sufficient (that is to
say, that the validity of the T-B law would be confirmed). Formally, it solves
the quantization problem completely in accord with numerical data [93]. The
problem lies however in the fact that these data were fitted to the power law,
Eq.(37), in accord with the original T-B empirical guess. Heisenberg's
honeycomb rule, Eq.(5), does \textsl{not} require the specific $n-$dependence.
\ In fact, we have to consider the observed (the Titius-Bode-type)
$n-$dependence only as a hint. With theoretical guidance emerging from this
work, it is hoped, that the attempts will be made to fit the observational
data to the Morse-like spectra \ in a way it is done routinely in chemical
physics for the Morse-type potentials. In this work we intentionally avoid use
of any adjustable parameters since the developed procedure, when supplied with
correctly interpreted numerical data, should be sufficient for obtaining
results without any adjustable parameters.

\ Having said this, we must notice that there is still room for improving
results we just obtained. Indeed, the constant $c$ was determined using the
harmonic approximation for the Morse-type potential. This approximation might
fail very quickly as the following arguments indicate. Although we can
calculate $r_{n}^{\prime}s$ using the T-B like law, Eq.(37), the arguments
following this equation cause us to use the equation $-\tilde{E}_{n}\equiv
-\ln\left\vert E\right\vert =-2\ln\sqrt{2}\pi+\ln r_{n\text{ \ }}$ for this
purpose. This means that we have to use Eq.(46) (with ground state energy
subtracted) in this equation in order to obtain the result for $r_{n}. $If we
ignore the quadratic correction in (46) (which is equivalent of calculating
the constant $c$ using harmonic oscillator approximation to the Morse
potential) then, by construction, we recover the T-B result, Eq.(46). If,
however, we do not resort to such an approximation, calculations will become
much more elaborate and are not physically illuminating. This is so because
the T-B law, Eq.(37), is a purely empirical best fit to the observed data. In
view of our calculations, Eq.(37) should be replaced by a more elaborate
fitting result \ which is in agreement with data for the Morse-type potential.
Since corrections to the harmonic oscillator potential in the case of the
Morse potential are typically small, they do not change things qualitatively.
Hence, we do not account for these complications in our paper. Nevertheless,
accounting for these (anharmonic) corrections readily explains why the
empirical T-B law works well for small n's and becomes increasingly unreliable
for larger n's [91].

In support of our conjectures we performed similar calculations for satellite
systems of Jupiter, Saturn, Uranus and Neptune. \ To do such calculations the
astronomical system of units is not immediately useful since in the case of
heavy planets one cannot use the relation $4\pi^{2}/\gamma M_{\odot}=1.$ This
is so because we have to replace the mass of the Sun $M_{\odot}$ by the mass
of the respective heavy planet. To do so, we write $4\pi^{2}=$\ $\gamma
M_{\odot}$, multiply both \ sides by $M_{j\text{ }}$\ (where $j$ stands for
the $j$-th heavy planet) and divide both sides by $M_{\odot}$. Thus, we
obtain: $4\pi^{2}q_{j}=$\ $\gamma M_{j}$,\ where $q_{j}=$\ $\dfrac{M_{j}%
}{M_{\odot}}$\ . Since the number $q_{j}$ is of order $10^{-3}$\ $-10^{-5}$,
it is inconvenient in actual calculations. To by pass this difficulty, we need
to readjust Eq.(40)\ by rescaling $x$ coordinate as $x=\delta\bar{x}$\ and, by
choosing \ $\delta^{2}$\ $=q_{j}$. After transition to quantum mechanics such
a rescaling results in \ replacing Eq.(42) for the spectrum by the following
result:%
\begin{equation}
-\tilde{E}_{n}=\frac{\gamma M}{2}[1-\frac{c\delta\ln\tilde{A}}{\sqrt{\gamma
M}}(n+\frac{1}{2})]^{2}. \tag{47}%
\end{equation}
Since the constant $c$\ is undetermined initially, we can replace it by
$\tilde{c}=c\delta$ so that \ we reobtain back equation almost identical to
Eq.(46). That is
\begin{equation}
-\tilde{E}_{n}=2\pi^{2}[1-\frac{(n+\frac{1}{2})}{4\pi^{2}}\ln\left(
\frac{\gamma M_{j}}{(r_{j})_{1}}\right)  ]^{2} \tag{48}%
\end{equation}
In this equation \ $\gamma M_{j}=$\ $4\pi^{2}q_{j}$ and \ $(r_{j})_{1}$ is the
semimajor axis of the satellite lying in the equatorial plane and closest to
the $j$-th planet. Our calculations are summarized in the Table 4 below.
Appendix C contains the input data used in our calculations of n$_{theory}%
^{\ast}.$\ \ \ \ \ \ \ \ \ \ \ \ \ \ \ \ \ \ \ \ \ \ \ \ \ \ \ \ \ \ \ \ \ \ \ \ \ \ 

\ \ \ \ \ \ \ \ \ \ \ \ \ \ \ \ \ \ \ \ \ \ \ \ \ \ \ \ \ \ \ \ \ \ \ \ \ \ \ \ \ Table
4%
\[%
\begin{tabular}
[c]{|l|l|l|}\hline
Satellite system%
$\backslash$%
n$_{\max}$ & n$_{theory}^{\ast}$ & n$_{\text{obs}}^{\ast}$\\\hline
Solar system & 9 & 9\\\hline
Jupiter system & 11-12 & 8\\\hline
Saturn system & 20 & 20\\\hline
Uranus system & 40 & 18\\\hline
Neptune system & 33 & 6\\\hline
\end{tabular}
\]
Since the discrepancies for Uranus and Neptune systems may be genuine or not
we come up with the following general pattern described below.

\subsection{Further analogies with atomic mechanics}

From atomic mechanics we know that the approximation of independent electrons
used by Bohr fails rather quickly with increased number of electrons. For this
reason to expect that the T-B law is going to hold for satellites of \ all
heavy planets is rather naive as we explained already. At the same time, for
planets rotating around the Sun such an approximation is seemingly good. The
SO(2,1) symmetry explains why the motion of all planets should be planar but
it does not explain why the motion of all planets is taking place in the plane
almost coinciding with the equatorial plane of the Sun. The same is true for
the regular satellites of all heavy planets as discussed by Dermott [95]. If
we adopt the quantum mechanical point of view, then we should accept that such
an arrangement of planets is the result of some kind of spin-orbital
interaction whose exact quantum mechanical nature remains to be elucidated.
Other rotational resonances ubiquitous in the Solar system could then be
explained quantum mechanically as well. The equatorial plane in which planets
(satellites) move can be considered as some kind of an orbital (in the atomic
physics terminology).\ It is being filled in accordance with the equivalent of
the Pauli principle:\textsl{\ each orbit can be occupied by no more than one
planet}\footnote{The meteorite belt can be looked upon as some kind of a ring.
We shall briefly discuss the rings below.}. Once the orbital is filled, other
orbitals associated with other planes will begin to be filled
out\footnote{Incidentally, such a requirement automatically excludes Pluto
from the status of a planet. Indeed, \ although the T-B-type law,
Eq.s(36),(37), can be seemingly adjusted to accomodate Pluto. Not only this
would contradict the data summarized in Table 4 but also would be in
contradiction with observational astronomical data for Pluto. According to
these data the orbit inclination for Pluto is 17$^{o}$ as compared to the rest
of planets whose inclination is within boundary margins of $\pm$2$^{o}($
except for Mercury for which it is 7$^{o}).$}. Some of orbitals can be empty.
This is indeed observed [95]. It should be said though that it appears
(according to available data, e.g. see [92], that not all of observed
satellites are moving on stable orbits. It appears\ also as if the "inner
shell", \ when completely filled, acts as some kind of an s-type spherical
orbital since \textsl{the} \textsl{orbits of} \textsl{other satellites lie
strictly outside the sphere whose diameter is greater or equal to that
corresponding to the last allowed energy level in the first shell}. The
location of secondary planes appears to be quite arbitrary as well as the
filling of their stable orbits. Furthermore, \ without account of spin-orbital
interactions, quantum mechanics says nothing about the direction of orbital
rotation. Although for all planets it does coincide with the direction of
rotation of Sun's axis, in the case of \ Phoebe- the irregular satellite of
Saturn-rotation takes place in the opposite direction to that of the axis of
Saturn. If the spin-orbital interaction does exists, most likely, Phoebe's
orbit is not a stable one.

It is tempting to extend the picture just sketched beyond the scope of our
Solar System. If for a moment we would ignore relativistic effects (they will
be discussed in the next section), we can then find out that our Sun is moving
along almost circular orbit around our galaxy center with the period
$T=185\cdot10^{6}$ years [98]. Our galaxy is also flat as our Solar System and
the major mass is concentrated in the galaxy center. Hence, again, \ if we
believe that stable stellar motion \ is taking place along the geodesics in
accordance with laws of Einstein's general relativity, then \ we have to
accept that our galaxy is a quantum object. It would be very interesting to
estimate the number of allowed energy levels for our galaxy and to check if
the Pauli-like principle works for the galaxy as well.

\subsection{Latest developments supporting our point of view}

\bigskip

We begin with the following observation. The motion of a planet of mass
$m_{0}$ in the field of two static centers of attraction with masses $m_{1}$
and $m_{2}$ was discussed by Legendre and Jacobi in 19th century [63] in
connection with their study of elliptic functions. Such an idealized problem
is a precursor of the restricted 3-body problem to be discussed in the next
subsection in connection with dynamics of planetary rings. As simple as it is,
the full study of this problem is extremely complex. It involves
classification of all points and lines of equilibria and motions in the
domains restricted by these lines. In addition to the eight major types of
\ bounded orbits there are many more coming from collision of equilibrium
point/lines etc. Characterization of the unbound motion is also interesting
but is less complex. In quantum mechanics the motion of an electron in the
presence of two fixed positive ions is also a benchmark problem (in addition
to study of He discussed in Section 1). Normally, charges of ions are assumed
to be the same (e.g. for H$_{2}^{+}$ ) which makes such a problem \ somewhat
different (since they repel each other) from the problem studied by Legendre
and Jacobi. All classification of molecular spectra can be traced back to this
problem [97]. As in \ the case of H atom, the correspondence principle is not
well established in this case since (to our knowledge) nobody studied the
agreement between the quantum -mechanical calculations in the semiclassical
limit and the results of Legendre-Jacobi theory modified due to the chemical
requirements. Interestingly enough such a comparison was made to a larger
extent between the classical restricted 3-body problem and its quantum analog.
The quantum analog of the restricted 3-body problem exists in the form of the
H atom placed in a strong crossed constant electric and magnetic fields [99].
Since semiclassical and classical analysis of such a system is sufficiently
well understood, this fact allows such a system to be studied both
theoretically and experimentally. These studies are well summarized in two
recent reviews [2,3] to which we refer our readers for details. For immediate
purposes of this work the following quotation from Porter and Cvitanovi\v{c}
is helpful: " almost perfect parallel between the governing equations of
atomic physics and celestial mechanics implies that the transport mechanism
for these two situations is virtually identical: on the celestial scale,
transport takes a spacecraft from one Lagrange point to another until it
reaches its desired destination\footnote{E.g. see also paper by Convay at al
[4].}. On the atomic scale, the same type of trajectory transports an electron
initially trapped near the atom across the escape threshold (in chemical
parlance, across a "transition state"), never to return. The orbits used to
design space missions thus also determine the ionization rates of atoms and
chemical reaction rates of molecules". This statement is nicely illustrated in
the paper by Jaffe et al [100] in which it is demonstrated that the
\textsl{transition state theory developed initially in chemistry} (to describe
the rates of chemical reactions) \textsl{is} \textsl{working actually better
in celestial mechanics} where the discrepancy between the chemical theory and
numerical simulations (done for celestial mechanics transport problems) is
less than 1\%. It should be noted though that the calculations were done at
the classical level only (that is for a very large quantum numbers). The
current status of transition state theory at the quantum and classical levels
in chemistry can be found in the recent book by Micha and Burghardt [101]..

\subsection{The restricted 3-body problem and planetary rings}

\ \ \ \ \ \ \ Although the literature on restricted 3-body problem is
huge,\ we would like to discuss this problem from the point of view of \ its
connection with general relativity and quantization of planetary orbits
\ along the lines advicated in this paper.

We begin with several remarks. First, the existence of ring systems for all
heavy planets is well documented [92]. Second, these ring systems are
interspersed with satellites of these heavy planets. Third, both rings and
satellites lie in the respective equatorial planes so that the satellites move
on stable orbits. From these observations it follows that: \ \ \ \ 

a) While each of heavy planets is moving along the geodesics around Sun, the
respective satellites are moving along the geodesics around \ 

\ \ \ \ respective planets;

b) The motion of these satellites is almost circular.

The restricted 3-body problem can be formulated now as follows. Given that the
rings are made of some kind of small objects whose masses can be
neglected\footnote{This approximation is known as Hill's problem/approximation
in the restricted 3-body problem [18].} as compared to masses of both the
satellite(s) and the particular heavy planet. Following previously
discussed\ ideas by Laplace, we can ignore mutual gravitational \ interaction
between these objects. Under such conditions we end up with the restricted
three-body problem of motion of a given piece of a ring (of zero mass) in the
presence of two bodies of masses $m_{1}$ and $m_{2}$ respectively. To simplify
matters, one usually assumes that the motion of these two masses takes place
on a circular orbit with respect to their center of mass. Complications
associated with the eccentricity of such a motion are discussed in the book by
Szebehely [102] and can be taken into account if needed. They will be ignored
nevertheless in our discussion since we shall assume that satellites of heavy
planets move on geodesics so that the center of mass coincides with the
position of a heavy planet anyway thus making our computational scheme
compatible with Einsteinian relativity. By assuming that ring pieces are
massless we also are making their motion compatible with requirements of
general relativity \ since whatever orbits they may have-these are geodesics.

Thus far only \ the motion of satellites in the equatorial planes (of
respective planets) was considered as stable (and, hence, quantizable). The
motion of ring pieces was not accounted thus far by these stable orbits. The
task now lies in showing that satellites lying inside the respective rings of
heavy planets \ are essential for stability of these rings motion and,
\ hence, they are making it quantizable.

For the sake of space, we would like only to provide a sketch of arguments
leading to such a conclusion. Our task is greatly simplified by the fact that
\ a very similar situation exists for 3-body system such as Moon, Earth and
Sun. \ Dynamics of such a system was studied very thoroughly by Hill whose
work played profound role in Poincar$e^{\prime^{\prime}}$ studies of celestial
mechanics \ [88]. Recently, Avron and Simon\textbf{\ [}103\textbf{]} adopted
Hill's ideas in order to develop a formal quantum mechanical treatment of the
Saturn rings. In this work we follow the original Hill's ideas concerning
dynamics of the Earth-Moon-Sun system. We claim that, when these ideas are
looked upon from the point of view of modern mathematics of exactly integrable
systems, they enable us to describe not only the Earth-Moon-Sun system but
also the dynamics of rings of heavy planets. These modern mathematical methods
enable us to find a place for the Hill's theory within general quantization
scheme discussed in previous sections.

\subsubsection{\bigskip Basics of the Hill's equation}

To avoid repetitions, we refer our readers to the books of Pars [63],
Chebotarev [98] and Brouwer and Clemence [104] for detailed and clear account
of the restricted 3-body problem and Hill's contributions to Lunar theory.
Here we only summarize the ideas behind Hill's ground breaking work.

In a nutshell his method of studying Lunar problem can be considered as
extremely sophisticated improvement of previously discussed Laplace and
Lagrange method. Unlike Laplace, Hill realized that \ both Sun and Earth are
surrounded by the rings of influence\footnote{Related to the so called Roche
limit [92].}. The same goes for all heavy planets. Each of these planets and
each satellite of such a planet will have its own domain of influence whose
actual width is controlled by the Jacobi integral of motion. For the sake of
argument, consider the Saturn as an example. It has Pan as its the innermost
satellite. Both the Saturn and Pan have their \ respective domains of
influence. Naturally, we have to look for the domain of influence for the
Saturn. Within such a domain let us consider a hypothetical closed Kepler-like
trajectory. \ Stability of such a \ Lagrangian trajectory is described by the
Hill equation\footnote{In fact, there will be the system of Hill's equations
in \ general [98]. This is so since the disturbance of trajectory is normally
decomposed into that which is perpendicular and that \ which is parallel to
the \ Kepler's trajectory at a given point. We shall avoid these complications
in our work.}. Since such an equation describes a wavy-type oscillations
around the presumably stable trajectory, the parameters describing such a
trajectory are used as an input (perhaps, with subsequent adjustment) in the
Hill equation given by
\begin{equation}
\frac{d^{2}x}{dt^{2}}+(q_{0}+2q_{1}\cos2t+2q_{2}\cos4t+\cdot\cdot\cdot)x=0.
\tag{49}%
\end{equation}
If we would ignore all terms except $q_{0}$ first, we would \ naively obtain:
$x_{0}(t)=A_{0}cos$($t\sqrt{q_{0}}+\varepsilon).$ This result describes
oscillations around the equilibrium position along the trajectory with the
constant $q_{0}$ carrying information about this trajectory and the amplitude
$A$ is expected to be larger or equal to the average distance between the
pieces of the ring. This naive picture gets very complicated at once should we
use the obtained result as an input into Eq.(49). In this case the following
equation is obtained
\begin{equation}
\frac{d^{2}x}{dt^{2}}+q_{0}x+A_{0}q_{1}\{\cos[t(\sqrt{q_{0}}+2)+\varepsilon
]+\cos[t(\sqrt{q_{0}}-2)-\varepsilon]\}=0 \tag{50}%
\end{equation}
whose solution will enable us to determine $q_{1}$and $A_{1}$ using the
appropriate boundary conditions. Unfortunately, since such a procedure should
be repeated \ infinitely many times, it is obviously impractical. Hill was
able to design much better method. \ Before discussing Hill's equation from
the perspective of modern mathematics, it is useful \ to recall the very basic
classical facts \ about this equation summarized in the book by Ince [105].
For this purpose, we shall assume that the solution of Eq.(49) can be
presented in the form
\begin{equation}
x(t)=e^{\alpha t}%
{\textstyle\sum\limits_{r=-\infty}^{\infty}}
b_{r}e^{irt}. \tag{51}%
\end{equation}
Substitution of this result into Eq.(49) leads to the following infinite
system of linear equations%
\begin{equation}
(\alpha+2ri)^{2}b_{r}+%
{\textstyle\sum\limits_{k=-\infty}^{\infty}}
q_{k}b_{r-k}=0,\text{ }r\in\mathbf{Z.} \tag{52}%
\end{equation}
As in finite case, obtaining of nontrivial solution requires the infinite
determinant $\Delta(\alpha)$ to be equal to zero. \ This problem can be looked
upon from two directions: either all constants $q_{k}$ are assigned and one is
interested in the bounded solution of Eq.(51) for $t\rightarrow\infty,$ or one
is interested in relationship between constants \ made in such a way that
$\alpha=0.$ In the last case it is important to know wether there is one or
more than one of such solutions available. Although answers can be found in
the book by Magnus and Winkler [106], we follow McKean and Moerbeke [107],
Trubowitz [108] and Moser [109].

For this purpose, we need to bring our notations in accord with those used in
these references. Thus, the Hill operator is defined now as $Q(q)=-\frac
{d^{2}}{dt^{2}}+q(t)$ with periodic potential $q(t)=q(t+1).$ Eq.(49) can now
be rewritten as
\begin{equation}
Q(q)x=\lambda x. \tag{53}%
\end{equation}
This presentation makes sense since $q_{0}$ in Eq.(49) plays a role of
$\lambda$ in Eq$.(53).$ Since this is the second order differential equation,
it has formally 2 solutions. These solutions depend upon boundary conditions.
For instance, for \textsl{periodic} solutions such that $x(t)=x(t+2)$ the
"spectrum" of Eq.(53) is discrete and is given by
\[
-\infty<\lambda_{0}<\lambda_{1}\leq\lambda_{2}<\lambda_{3}\leq\lambda
_{4}<\cdot\cdot\cdot\uparrow+\infty.
\]
We wrote the word spectrum in quotation marks because of the following.
Eq.(53) does have a normalizable solution only if $\lambda$ belongs to the
(pre assigned) intervals $(\lambda_{0},\lambda_{1}),(\lambda_{2},\lambda
_{3}),...,(\lambda_{2i},\lambda_{2i+1}),...$ \ \ In such a case the
eigenfunctions $x_{i}$ are normalizable in the usual sense of quantum
mechanics and form an orthogonal set. The periodic solutions make sense for
vertical displacement from the reference trajectory. For the horizontal
displacement the boundary condition should be chosen as $x(0)=x(1)=0.$ For
such chosen boundary condition the discrete spectrum also exists but it lies
exactly in the gaps between the intervals just described, i.e. $\lambda
_{1}\leq\mu_{1}\leq\lambda_{2}<\lambda_{3}\leq\mu_{2}\leq\lambda_{4}%
<\cdot\cdot\cdot.$ For such a spectrum there is also set of normalized
mutually orthogonal eigenfunctions. Thus in both cases quantum mechanical
description is assured. One can do much more however. In particular, Trubowitz
[108] designed an explicit procedure of recovering the potential $q(t)$ from
the $\mu-$spectrum supplemented by information about normalization constants.

\ The Hill's equation can be interpreted in terms of the auxiliary dynamical
(Neumann) problem. Such an interpretation is very helpful for us since it
allows us to include the quantum mechanics of Hill's equation into general
formalism developed in Sections 2 and 3.

\subsubsection{Connection with the dynamical Neumann problem and the Korteweg
-de Vries equation}

Before describing such connenctions, we would like to add few details to the
results of previous subsection. First, the number of the pre assigned
intervals is always finite. This means that, beginning with some pre assigned
$\hat{\imath}$, we would be left with $\lambda_{2i}=\lambda_{2i+1}\forall
i>\hat{\imath}.$These \textsl{double} eigenvalues do not have independent
physical significance since they can be determined by the set of
\textsl{single} eigenvalues (for which $\lambda_{2i}\neq\lambda_{2i+1})$ as
demonstrated by Hochstadt [110]. Because of this, the potentials $q(t)$ in the
Hill's equation are called the \textsl{finite gap} potentials\footnote{Since
there is only finite number of gaps [$\lambda_{1},\lambda_{2}],$[$\lambda
_{3},\lambda_{4}],...$where the spectrum is forbidden.}. Hence, physically, it
is sufficient to discuss only such potentials which possess finite single
spectrum. The auxiliary $\mu-$spectrum is then determined by the gaps of the
single spectrum as explained above. With this information in our hands, we are
ready to discuss the exactly solvable Neumann dynamical problem. It is the
problem about dynamics of a particle moving on the $n-$dimensional sphere
$<\mathbf{\xi},\mathbf{\xi}>\equiv\xi_{1}^{2}$ +$\cdot\cdot\cdot+\xi_{n}%
^{2}=1$ under the influence of a quadratic potential $\phi(\mathbf{\xi
})=<\mathbf{\xi},\mathbf{A\xi}>.$ Equations of motion describing the motion on
$n-$ sphere are given by
\begin{equation}
\mathbf{\ddot{\xi}}=-\mathbf{A\xi}+u(\mathbf{\xi})\mathbf{\xi}\text{ \ with
}u(\mathbf{\xi})=\phi(\mathbf{\xi})-<\mathbf{\dot{\xi}},\mathbf{\dot{\xi}}>.
\tag{54}%
\end{equation}
Without loss of generality, we assume that the matrix $\mathbf{A}$ is already
in the diagonal form: $\mathbf{A}:=diag(\alpha_{1},...,\alpha_{n}). $ With
such an assumption we can equivalently rewrite (54) in the following
suggestive form%
\begin{equation}
\left(  -\frac{d^{2}}{dt^{2}}+u(\mathbf{\xi}(t))\right)  \xi_{k}=\alpha_{k}%
\xi_{k}\text{ ; \ }k=1,...,n. \tag{55}%
\end{equation}
Thus, in the case if we can prove that $u(\mathbf{\xi}(t))$ in (55) is the
same as $q(t)$ in (53), the connection between the Hill and Neumann's problems
will be established. The proof is presented in Appendix D. It is different
from that given in the lectures by Moser [109] since it is more direct and
much shorter.

This proof brought us the unexpected connection with hydrodynamics through
\ the static version of Korteweg-de Vries equation. Attempts to describe the
Saturnian rings using equations of hydrodynamic are described in the recent
monograph by Esposito [111]. This time, however, we can accomplish more using
\ just obtained information. This is the subject of the next subsection.

\subsubsection{Connections with SO(2,1) group and the K-Z equations}

\bigskip

Following Kirillov [112], we introduce the commutator for the fields
(operators) $\xi$ and $\eta$ as follows: [$\xi,\eta]=\xi\partial\eta
-\eta\partial\xi.$ Using the KdV, Eq.(D.10), let us consider 3 of its
independent solutions: $\xi_{0},\xi_{-1}$ and $\xi_{1}.$ All these solutions
can be obtained from general result: $\xi_{k}=t^{k+1}+O(t^{2}),$ valid near
zero. Consider now a commutator $[\xi_{0},\xi_{1}].$ Straightforwardly, we
obtain, $[\xi_{0},\xi_{1}]=\xi_{1}$. Analogously, we obtain, $[\xi_{0}%
,\xi_{-1}]=-\xi_{-1}$ and, finally, $[\xi_{1},\xi_{-1}]=-2\xi_{0}.$ According
to Kirillov, such a Lie algebra is isomorphic to that for the group $SL(2,R)$.
Vilenkin [113] demonstrated that the group $SL(2,R)$ is isomorphic to
$SU(1,1)$. Indeed, by means of transformation: \textit{w}\textsl{=}%
$\dfrac{\mathit{z}-i}{\mathit{z}+i},$ it is possible to transform the upper
half plane (on which $SL(2,R)$ acts) into the interior of unit circle on which
$SU(1,1)$ acts. Since, according to Appendix B, the group $SU(1,1)$ is the
connected component of $SO(2,1)$, the anticipated connection with $SO(2,1)$
group is established.

In Appendix D we noticed connections between the Picard-Fuchs, Hill and
Neumann-type equations. In a recent paper by Veselov et al [114] such a
connection was developed much further resulting in the K-Z type
equations\footnote{E.g. see Eq.(29) of Section 2.} for Neumann-type dynamical
systems. We refer our readers to the original literature, especially to the
well written lecture notes by Moser [109]. These notes as well and his notes
in collaboration with Zehnder [84] provide an excellent background for the
whole circle of ideas relating Hill's equation to integrable models.

\section{Solar System at larger scales: de Sitter, anti -de Sitter and
conformal symmetries compatible with orbital quantization}

\bigskip

The obtained results demonstrate a remarkable interplay between the Newtonian
and Einsteinian mechanics already at the scale of our Solar System. Since
quantization of stable orbits described in this paper is possible only with
use of \ the basic experimental facts assuring \ correctness of \ results of
general relativity, it is only natural to reverse this statement and to say
that \textsl{the correctness of general relativity is assured by the observed
pattern of stable (quantum) orbits}.

Since quantum mechanics can be developed group-theoretically, the same should
be true for relativity. Quoting Einstein, Infeld and Hoffmann [89]: "Actually,
the only equations of gravitation which follow without ambiguity from the
fundamental assumptions of the general theory of relativity are the equations
for empty space\footnote{See also Section 6.}, and it is important to know
whether they alone are capable of determining the motion of bodies". In this
work we argue that this is certainly correct locally when the Lorentzian-type
symmetry holds true. Now we would like to discuss how such locally Lorentzian
space-time embeds into space-times of general relativity possessing larger
symmetry groups\footnote{At the level of quantum field theory Utiyama [115]
demonstrated that the requirement of the local gauge invariance implemented
for the non Abelian Lorentz group produces the Einstein field equations for
gravitational field. This result implies that any "improvements" of
Einsteinian relativity should involve changes in the local Lorentzian
structure of space-time which is very unlikely.\ Independent arguments
supporting this point of view are presented in Section 6.}. Since this topic
is extremely large, we shall discuss only the most basic facts from the point
of view of results obtained in this paper.

To our knowledge, Dirac [116] was the first who recognized the role of
space-time symmetry in quantum mechanics. In his paper he wrote: "The
equations of atomic physics are usually formulated in terms of space-time of
special relativity. They then have to form a scheme which remains invariant
under all transformations which carry the space-time over into itself. These
transformations consist of the Lorentz rotations about a point combined with
arbitrary translations, and form a group.... Nearly all of more general spaces
have only trivial groups\footnote{This statement of Dirac is not correct.
However, it is correct at the time of writing of his paper.}of operations
which carry the spaces into themselves....There is one exception, however,
namely the de Sitter space (with no local gravitational fields). This space is
associated with a very interesting group, and so the study of the equations of
atomic physics in this space is of special interest, \textsl{from mathematical
point of view}." Subsequent studies indicated that the symmetry of space-time
could be important even at the atomic scale [117,118]. Another reason to look
at larger symmetry groups is associated with the cosmological constant problem
[119] and, \ the associated with it problem of existence of cold dark energy
(CDE) [120] cold dark matter (CDM) [121] and the modified Newtonian dynamics
(MOND) [122]. Clearly, we are unable to discuss these issues within the scope
of this paper since they are more relevant to processes at galactic scales.
Nevertheless, we would like to notice that, for instance, the MOND presupposes
use of Newtonian and the modified Newtonian mechanics at the galactic scales
which, as discussed in Section 4, strictly speaking, \ is not permissible even
at the scales of our Solar System. The rationale for the dark energy and dark
matter is explained in our recent paper [123] based on mathematical arguments
consistent with that used by Grigory Perelman in his proof of the
Poincare$^{\prime}$ conjecture.

Hence, we proceed with description of the de Sitter and anti-de Sitter spaces
based on results of our recent work. We begin with the following
\ Hilbert-Einstein functional
\begin{equation}
S^{c}(g)=%
{\textstyle\int\nolimits_{\mathcal{M}}}
d^{d}xR\sqrt{g}+\Lambda%
{\textstyle\int\nolimits_{\mathcal{M}}}
d^{d}x\sqrt{g} \tag{56}%
\end{equation}
defined for some (pseudo) Riemannian manifold $\mathcal{M}$ of total
space-time dimension $d$. The (cosmological) constant $\Lambda$ is determined
as follows.

Using $R_{ij}$, the Ricci curvature tensor, the \textsl{Einstein space} is
defined as solution of the following vacuum Einstein equation%
\begin{equation}
R_{ij}=\lambda g_{ij} \tag{57}%
\end{equation}
with $\lambda$ being a constant. From this definition it follows that
\begin{equation}
R=d\lambda. \tag{58}%
\end{equation}
At the same time, variation of the action $S^{c}(g)$ produces%
\begin{equation}
G_{ij}+\frac{1}{2}\Lambda g_{ij}=0, \tag{59}%
\end{equation}
where the Einstein tensor $G_{ij}$ is defined as $G_{ij}=R_{ij}-\frac{1}%
{2}g_{ij}R$ with $R$ being the scalar curvature determined by the metric
tensor $g_{ij}\footnote{Eq.(59) illustrates the meaning of the term "dark
matter". The constant $\Lambda$ enters into the stress-energy tensor (in the
present case given by $-$ $\frac{1}{2}\Lambda g_{ij})$ typically associated
with the matter, Einstein [82].}.$ Combined use of Eq.s(58) and (59) produces:
$\Lambda=\lambda(d-2).$ Substitution of this result back into Eq.(59)
produces:%
\begin{equation}
G_{j}^{i}=(\frac{1}{d}-\frac{1}{2})\delta_{j}^{i}R. \tag{60}%
\end{equation}
Since\ by design $G_j,h^i=0,$ \ we obtain our major result:
\begin{equation}
(\frac{1}{d}-\frac{1}{2})R_{,j}=0, \tag{61}%
\end{equation}
implying that scalar curvature $R$ is constant.

For isotropic homogenous spaces the Riemann curvature tensor can be presented
in the following known form [81]:
\begin{equation}
R_{ijkl}=k(x)(g_{ik}g_{jl}-g_{il}g_{jk}). \tag{62}%
\end{equation}
Accordingly, the Ricci tensor is obtained as: $R_{ij}=k(x)g_{ij}(d-1).$ The
Schur's theorem [124] guarantees that for $d\geq3$ we must have $k(x)=k=const$
for the entire space. Therefore, we obtain: $\lambda=(d-1)k$ and, furthermore,
$R=d(d-1)k.$ The spatial coordinates can always be rescaled so that $R=k$ or,
alternatively, the constant $k$ can be normalized to unity. For $k>0,$ $k=0$
and $k<0$ we obtain respectively de Sitter, flat and anti-de Sitter spaces.
Thus, we just have demonstrated that homogeneity and isotropy of space-time is
synonymous with spaces being de Sitter, flat and anti-de Sitter very much like
in ordinary Riemannian geometry there are spaces of positive, negative and
zero curvature. This fact can be used to give the alternative description of
\ just obtained results.

We begin with simple observation that the surface of \ constant positive
curvature is conformally equivalent to a sphere embedded in the Euclidean
space [123]. In particular, let us consider a 3-sphere embedded into 4d
Euclidean space. It is described by the equation
\begin{equation}
S^{3}=\{x\in E_{4},\text{ }x_{1}^{2}+x_{2}^{2}+x_{3}^{2}+x_{4}^{2}=R^{2}\}.
\tag{63}%
\end{equation}
$S^{3}$ is homogenous isotropic space with \ positive scalar curvature whose
value is $6/R^{2}.$ The group of motions associated with this homogenous space
is the rotation group $SO(4)$. The space of constant negative curvature
$H^{3}$ is obtained analogously. For this purpose it is sufficient, following
Dirac [116], to make $x_{1}$ purely imaginary and to replace $R^{2}$ by
$-R^{2}$ in Eq.(63). Such replacements produce:
\begin{equation}
H^{3}=\{x\in M_{4},\text{ }x_{1}^{2}-x_{2}^{2}-x_{3}^{2}-x_{4}^{2}=R^{2}\}.
\tag{64}%
\end{equation}
In writing this result we have replaced the Euclidean space $E_{4}$ by the
Minkowski space $M_{4}$ so that the rotation group $SO(4)$ is now replaced by
the Lorentz group $SO(3,1)$. The de Sitter space can now be obtained according
to Dirac (1935) as follows. In Eq.(63) we replace $E_{4}$ by $E_{5} $ and make
$x_{1}$ purely imaginary \ thus converting $E_{5}$ into $M_{5}$. The obtained
space is the de Sitter space whose group of symmetry is $SO(4,1) $%
\begin{equation}
dS_{4}=\{x\in M_{5},\text{ }x_{1}^{2}-x_{2}^{2}-x_{3}^{2}-x_{4}^{2}-x_{5}%
^{2}=R^{2}\}. \tag{65}%
\end{equation}
It has a constant positive scalar curvature whose value is $12/R^{2}.$ Very
nice description of such a space is contained in the book by Hawking and Ellis
[125]. The connection between parameter $R$ and the cosmological constant
$\Lambda$ is given by $R=\sqrt{\dfrac{3}{\Lambda}}$. The anti-de Sitter space
is determined analogously as also discussed by Hawking and Ellis and by Dirac.
Specifically, it is given by
\begin{equation}
adS_{4}=\{x\in E_{3,2},\text{ }x_{1}^{2}-x_{2}^{2}-x_{3}^{2}-x_{4}^{2}%
+x_{5}^{2}=R^{2}\}, \tag{66}%
\end{equation}
where the five dimensional space $E_{3,2}$ is constructed by adding the
time-like direction to $M_{4}.$ Hence, the symmetry group of $adS_{4}$ is
SO(3,2). All these groups can be described simultaneously if, following Dirac
[116], we introduce the quadratic form
\begin{equation}%
{\textstyle\sum\limits_{\mu=1}^{5}}
x_{\mu}x_{\mu}=R^{2} \tag{67}%
\end{equation}
in which some of the arguments are allowed to be purely imaginary.
Transformations preserving such a quadratic form are appropriate respectively
for groups SO(5), SO(4,1) and SO(3,2). \ We still can embed all these groups
into a larger (conformal) group SO(4,2) by increasing summation from 5 to 6 in
Eq.(67). In such a case all groups discussed in this work, starting from
SO(2,1), can be embedded into this conformal group as subgroups as discussed
in great detail by Wybourne [70]\footnote{Incidentally, the work by Graner and
Dubrulle [126], when \ translated into group-theoretic language, becomes
\ just \ a corollary of conformal invariance implied by the conformal group
.}. Comprehensive group-theoretic description of the Einstein spaces, e.g. see
Eq.(57), including those which are invariant with respect to the conformal
group, can be found in the monograph by Petrov [127]. The significance and use
of conformal symmetry in both gravity and conformal field theories has been
recently extended in [123]. All existing cosmological models \ in the limit
$R\rightarrow\infty$ approach one of the Einstein's spaces whose group of
symmetry belongs to the types just described. The de Sitter and anti-de Sitter
spaces are the simplest examples of such spaces [128].\footnote{The most
recent mathematically rigorous description of both de Sitter and anti-de
Sitter spaces can be found in the paper by Andersson and collaborators [129].}.

The task still remains to find out if representations of these larger groups,
e.g. see Vilenkin [113] and Wybourne [70] for mathematical details, \ can
\ produce the exact solutions of \ radial Schr\"{o}dinger equations not listed
in the Natanzon-style classification, e.g. see Levai [72], for SO(2,1). If
such solutions do exist, one might be able to find those of them which are of
relevance to celestial quantum mechanics and, hence, to cosmology.

\section{Conclusions}

\ It is a remarkable historical fact (discussed in Section 4.2.) that Laplace
was the first who studied resonance dynamics of known satellites of Jupiter
(effectively) using geodesics while Lagrange analyzing motion of these
satellites along stable geodesics had arrived (effectively) at the
Bohr-Sommerfeld quantization condition. The validity of the geodesic-type
approximation is based on the equivalence principle of general relativity
correct in the limit of vanishingly small masses as compared to the mass of
the central body. Using this principle Einstein [82] was able to calculate the
perihelium shift for Mercury. Within approximations he made all interactions
of Mercury with the rest of planets were ignored. Because of this, the same
type of calculations in the spirit of Laplace can be made for all planets as
discussed by Misner et al [83]. The fundamental problem then lies in proving
that such stable motions will survive in the case if masses of planets are
small but nonzero. In the case of satellites of Jupiter such a task was
completed by Poincare$^{\prime}$ and de Sitter as mentioned in Section 4. The
latest advancements are also discussed in the same section.

In this (concluding) section we argue that it is possible to arrive at the
same field equations of general relativity \textsl{by entirely by passing the
equivalence principle.} This can be achieved by studying the limiting case of
dynamics of 2+1 gravity as discussed in our papers [130-132] and in more
recent paper by 't Hooft [133]. In such a limit one studies surface dynamics
of the fictitious 2 dimensional gravitating bodies\footnote{These can be
actually visualized by the crossections (lying on the surface) of infinitely
long and thin massive 3 dimensional rods. According to the imposed rules we
are allowed to watch only the motion of crossections.}. Simple topological
arguments applied to this case indicate that the Einstein field equations
survive such a reduction while Newtionian (actually, the Poissonian-type)
equations \textsl{do not} survive this reduction. As result, \textsl{the
dynamics of 2+1 gravity\ is strictly} \textsl{Einsteinian}. In mathematics
this type of dynamics ( the dynamics of measured foliations) was discovered
totally independently of gravity-related considerations by Thurston [134] in
his study of 3-manifolds\footnote{He was awarded the Fields medal for these
studies.}. Physically, such type of dynamics is realized in dynamics of some 2
dimensional liquid crystals. All this is explained in \ our works [130,131]
which, in turn, were inspired by the earlier work by Deser\textbf{,} Jackiw
and t'Hooft [135]. The theory of foliations (for surfaces) is thoroughly
discussed in monograph by Nikolaev [136]. The book by Hehl\textbf{\ }and
Obukhov [137] uses foliations for description of classical electrodynamics in
3+1 space. Some basics of foliation theory from the point of view of Lie
groups and Lie algebras are discussed in the very readable book by Moerijk and
Mr\v{c}un [138]. By reversing reduction arguments it is possible to arrive at
dynamics of full 3+1 gravity in the presence of matter beginning from the
dynamics of measured foliations for 2+1 gravity. The \ topological properties
of dynamics of 2+1 gravity are described in terms of \ polynomials of knots
and links as explained in detail in the paper by Kholodenko [132].
Accordingly, dynamics of full 3+1 gravity should be associated with time
dynamics of 3-manifolds foliating 3+1 space [123]. These remarks provide
needed physical justification to works by Witten [138] and Kholodenko [139]
connecting statics and dynamics of 3 (or 2+1) gravity with conformal and
string theories.

\ 

\textbf{Acknowledgements }\ The Tables 1 and 2 are reproduced with permission
of the Copyright Clearance Center, Inc.(222 Rosewood Drive, Danvers, Ma
01923). The author would like to thank Dr.Jack Douglas (NIST) for many useful
comments and references.

\ \ 

\textbf{Appendix A. Details of Heisenberg's derivation of the commutator
identity }$[\hat{x},\hat{p}]=i\hbar$

\bigskip

In this appendix we would like to provide some details of Heisenberg's
reasoning leading to the discovery of $[\hat{x},\hat{p}]=i\hbar$. This would
be unnecessary should his original paper [27] contain all details.

At the classical level consider a gas of noninteracting atoms, better just one
atom containing $N$ electrons which are assumed to scatter light
independently. The interaction between the incoming light and individual
electron is described with help of the combination \textbf{d}$=\beta$\textbf{E
}where\textbf{\ d }is the dipole moment of the electron in the atom,
\textbf{E} is the strength of the external electric field which is assumed to
be time-dependent, and $\beta$ is the polarization tensor (in the simplest
case it is assumed to be a scalar). In the medium the strength of the electric
field changes as compared to the vacuum. By denoting it as \textbf{D} it is
known that \textbf{D}=\textbf{E} +4$\pi\mathbf{P}$ where \textbf{P}%
=N\textbf{d. }Since, at the same time, by definition, \textbf{d}=e\textbf{r
}we have to have an equation for\textbf{\ r. }It is given by%
\begin{equation}
\mathbf{\ddot{r}}+\omega_{0}^{2}\mathbf{r+\gamma\dot{r}=}\frac{e}%
{m}\mathbf{E(}t\mathbf{)} \tag{A.1}%
\end{equation}
where $e$ is electron's charge and $m$ is its mass. In writing this equation
it is assumed that our electron is bound harmonically (with the basic
frequency $\omega_{0}^{2})$ and that the friction is of \ known
(electromagnetic) nature and is assumed to be small. Using Fourier
decomposition of \textbf{r}(t) we obtain,%
\begin{equation}
\mathbf{r}(\omega)=\frac{e}{m}\frac{\mathbf{E}}{\omega_{0}^{2}-\omega
^{2}+i\omega\gamma}. \tag{A.2}%
\end{equation}
This equation allows\ us to obtain \textbf{P} and, hence, \textbf{D }as
follows\textbf{\ }:%
\begin{equation}
\mathbf{D}=\mathbf{E}+4\pi\mathbf{P}=(1+4\pi N\text{ }\frac{e^{2}}{m}\frac
{1}{\omega_{0}^{2}-\omega^{2}+i\omega\gamma})\mathbf{E}\equiv\varepsilon
(\omega)\mathbf{E.} \tag{A.3}%
\end{equation}
This equation defines a complex frequency-dependent dielectric constant
$\varepsilon(\omega).$ From electrodynamics it can be equivalently rewritten
as $\varepsilon(\omega)=(n(\omega)-i\varkappa(\omega))^{2}$ where $n(\omega)$
is the refractive index while $\varkappa(\omega)$ is the coefficient of
absorption. Using these facts we can write approximately%
\begin{equation}
n(\omega)=1+2\pi N\text{ }\frac{e^{2}}{m}\frac{1}{\omega_{0}^{2}-\omega
^{2}+i\omega\gamma}. \tag{A.4}%
\end{equation}
By ignoring friction in the high frequency limit we obtain,%
\begin{equation}
n(\omega)=1-2\pi N\text{ }\frac{e^{2}}{m\omega^{2}}. \tag{A.5.}%
\end{equation}
To account for quantum mechanical effects, Thomas, Reich and Kuhn in 1925
(just \textsl{before} the quantum mechanics was born !) have suggested to
replace Eq.(A.4) by
\begin{equation}
n(\omega)=1+2\pi N\text{ }\frac{e^{2}}{m}\sum\limits_{i}\frac{f_{i}}%
{\omega_{i0}^{2}-\omega^{2}} \tag{A.6}%
\end{equation}
where, following these authors, we ignored friction and introduced the
\textit{oscillator strength} $f_{i}.$To reconcile Eq.(A.6) with (A.5) we have
to require $\sum\limits_{i}f_{i}=1.$ This requirement is known as the
\textit{sum rule}. \ These facts were known to Kramers and
Heisenberg\footnote{E.g. see the reference in Heisenberg's paper.} where our
readers can find additional details. To make our point and to save space, we
would like to reobtain the result, Eq.(A.6), quantum mechanically \ using
modern formalism. We refer our reader to the book by Davydov\textbf{\ [}%
140\textbf{] } for additional details. Basically, we need to calculate quantum
mechanically the dipole moment \textbf{d, }that is
\begin{equation}
\mathbf{d}_{m}=\int\psi_{m}^{\ast}e\mathbf{r}\psi_{m}d^{3}\mathbf{r.}
\tag{A.7.}%
\end{equation}
In this expression the wave function $\psi_{m}$ is calculated with help of the
stationary perturbation theory with accuracy up to the first order in
perturbation (which is e\textbf{r}$\cdot$\textbf{E}). A short calculation
produces the following result for the oscillator strength,%
\begin{equation}
f_{km}=\frac{2m\omega_{km}}{\hbar}\left\vert \left\langle k\mid\hat{x}\mid
m\right\rangle \right\vert ^{2}. \tag{A.8}%
\end{equation}
This result can be equivalently rewritten as
\begin{equation}
f_{km}=\frac{m\omega_{km}}{\hbar}\{\left\langle k\mid\hat{x}\mid
m\right\rangle ^{\ast}\left\langle k\mid\hat{x}\mid m\right\rangle
+\left\langle k\mid\hat{x}\mid m\right\rangle ^{\ast}\left\langle k\mid\hat
{x}\mid m\right\rangle \}. \tag{A.9}%
\end{equation}
Since, however,%
\begin{equation}
im\omega_{km}\left\langle k\mid\hat{x}\mid m\right\rangle =\left\langle
k\mid\hat{p}_{x}\mid m\right\rangle \tag{A.10}%
\end{equation}
we can rewrite Eq.(A.9) as
\begin{equation}
f_{km}=\frac{1}{i\hbar}\{\left\langle m\mid\hat{x}\mid k\right\rangle
\left\langle k\mid\hat{p}_{x}\mid m\right\rangle -\left\langle m\mid\hat
{p}_{x}\mid k\right\rangle \left\langle k\mid\hat{x}\mid m\right\rangle \}
\tag{A.11}%
\end{equation}
since $\omega_{km}=-\omega_{mk}$. Finally, we have to require $\sum
\limits_{k}f_{km}=1.$ This is possible only if
\begin{equation}
\frac{1}{i\hbar}\left\langle m\mid\hat{x}\hat{p}_{x}-p_{x}\hat{x}\mid
m\right\rangle =1, \tag{A.12}%
\end{equation}

QED.

\ 

\textbf{Appendix B. Some quantum mechanical problems associated with the Lie
algebra of} \textbf{SO(2,1)} \textbf{group}\bigskip

Following Wybourne [70] let us consider the second order differential equation
of the type%
\begin{equation}
\frac{d^{2}Y}{dx^{2}}+V(x)Y(x)=0 \tag{B.1}%
\end{equation}
where $V(x)=a/x^{2}+bx^{2}+c.$ Consider as well the Lie algebra of the
noncompact group SO(2,1) or, better, its connected component SU(1,1). It is
given by the following commutation relations%
\begin{equation}
\lbrack X_{1},X_{2}]=-iX_{3};\text{ }[X_{2},X_{3}]=iX_{1};\text{ }[X_{3}%
,X_{1}]=iX_{2} \tag{B.2}%
\end{equation}
We shall seek the realization of this Lie algebra in terms of the following
generators%
\begin{equation}
X_{1}:=\frac{d^{2}}{dx^{2}}+a_{1}(x);\text{ \ }X_{2}:=i[k(x)\frac{d}{dx}%
+a_{2}(x)];\text{ \ }X_{3}:=\frac{d^{2}}{dx^{2}}+a_{3}(x). \tag{B.3}%
\end{equation}
The unknown functions $a_{1}(x),a_{2}(x),a_{3}(x)$ and $k(x)$ are determined
upon substitution of Eq.s(B.3) into Eq.s(B.2). After some calculations, the
following result is obtained%
\begin{equation}
X_{1}:=\frac{d^{2}}{dx^{2}}+\frac{a}{x^{2}}+\frac{x^{2}}{16};\text{ }%
X_{2}:=\frac{-i}{2}[x\frac{d}{dx}+\frac{1}{2}];\text{ }X_{3}:=\frac{d^{2}%
}{dx^{2}}+\frac{a}{x^{2}}-\frac{x^{2}}{16}. \tag{B.4}%
\end{equation}
In view of this, Eq.(B.1) can be rewritten as follows
\begin{equation}
\lbrack(\frac{1}{2}+8b)X_{1}+(\frac{1}{2}-8b)X_{3}+c]Y(x)=0 \tag{B.5}%
\end{equation}
This expression can be further simplified by the unitary transformation$UX_{1}%
U^{-1}=X_{1}\cosh\theta+X_{3}\sinh\theta;$ $UX_{3}U^{-1}=X_{1}\sinh
\theta+X_{3}\cosh\theta$ with $U=exp(-i\theta X_{2}).$ By choosing
$\tanh\theta=-(1/2+8b)/(1/2-8b)$ Eq.(B.5) is reduced to
\begin{equation}
X_{3}\tilde{Y}(x)=\frac{c}{4\sqrt{-b}}\tilde{Y}(x) \tag{B.6}%
\end{equation}
where the eigenfunction $\tilde{Y}(x)=UY(x)$ is an eigenfunction of both
$X_{3}$ and the Casimir operator \textbf{X}$^{2}=X_{3}^{2}-X_{2}^{2}-X_{1}^{2}
$ so that by analogy with the Lie algebra of the angular momentum we obtain,
\begin{align}
\mathbf{X}^{2}\tilde{Y}_{jn}(x)  &  =J(J+1)\tilde{Y}_{Jn}(x)\text{
\ \ and}\tag{B.7a}\\
X_{3}\tilde{Y}_{Jn}(x)  &  =\frac{c}{4\sqrt{-b}}\tilde{Y}_{Jn}(x)\equiv
(-J+n)\tilde{Y}_{Jn}(x)\text{; }\ n=0,1,2,...\text{.} \tag{B7b}%
\end{align}
It can be shown that $J(J+1)=-a/4-3/16$. From here we obtain : $J=-\frac{1}%
{2}(1\pm\sqrt{\frac{1}{4}-a});$ $\frac{1}{4}-a\geq0.$ In the case of discrete
spectrum one should choose the plus sign in the expression for $J$. Using this
result in Eq.(B.7) we obtain the following result of major importance%
\begin{equation}
4n+2+\sqrt{1-4a}=\frac{c}{\sqrt{-b}}. \tag{B.8}%
\end{equation}
Consider now the planar Kepler problem. In this case, in view of Eq.(32), the
radial Schr\"{o}dinger equation can be written in the following symbolic form%
\begin{equation}
\left[  \frac{d^{2}}{dr^{2}}+\frac{1}{r}\frac{d}{dr}+\frac{\mathit{\upsilon}%
}{r}+\frac{u}{r^{2}}+g\right]  R(r)=0 \tag{B.9}%
\end{equation}
By writing $r=x^{2}$ and $R(r)=x^{-\frac{1}{2}}\mathcal{R}(x)$ This equation
is reduced to the canonical form given by Eq(B.1), e.g. to%
\begin{equation}
(\frac{d^{2}}{dx^{2}}+\frac{4u+1/4}{x^{2}}+4gx^{2}+4\upsilon)\mathcal{R}(x)=0
\tag{B.10}%
\end{equation}
so that the rest of arguments go through. Analogously, in the case of
Morse-type potential we have the following Schrodinger-type equation
initially:%
\begin{equation}
\left[  \frac{d^{2}}{dz^{2}}+pe^{2\alpha z}+qe^{\alpha z}+k\right]  R(z)=0
\tag{B.12}%
\end{equation}
By choosing $z=lnx^{2}$ and $R(z)=x^{-\frac{1}{2}}\mathcal{R}(x)$ Eq.(B12) is
reduced to the canonical form%
\begin{equation}
(\frac{d^{2}}{dx^{2}}+\frac{16k+\alpha^{2}}{4\alpha^{2}x^{2}}+\frac{4p}%
{\alpha^{2}}x^{2}+\frac{4q}{\alpha^{2}})\mathcal{R}(x)=0 \tag{B.13}%
\end{equation}
By analogous manipulations one can reduce to the canonical form the radial
equations for Hydrogen atom and for the 3-dimensional harmonic oscillator.

\bigskip

\textbf{Appendix C}. \textbf{Numerical data used for claculations of}
\textbf{n}$_{theory}^{\ast}$

(\textbf{Table 4}).

\ 

1 au=149.598$\cdot10^{6}km$

Masses (in kg): Sun 1.988$\cdot10^{30},$ Jupiter 1.8986$\cdot10^{27}$, Saturn
5.6846$\cdot10^{26},$

Uranus 8.6832$\cdot10^{25},$ Neptune 10.243$\cdot10^{25}.$

q$_{j}:$ Jupiter 0.955$\cdot10^{-3},$ Saturn 2.86$\cdot10^{-4},$ Uranus
4.37$\cdot10^{-5}$, Neptune 5.15$\cdot10^{-5}.$

$\left(  r_{j}\right)  _{1}(km):$ Jupiter 127.69$\cdot10^{3},$ Saturn
133.58$\cdot10^{3},$ Uranus 49.77$\cdot10^{3},$

Neptune 48.23$\cdot10^{3}.$

ln$\left(  \dfrac{\gamma M}{2r_{1}}\right)  $ : Earth 4.0062, Jupiter 3.095,
Saturn 1.844, Uranus 0.9513,

Neptune 1.15.

\ 

\textbf{Appendix D}. \textbf{Connections between the Hill and Neumann's }

\textbf{dynamical problems.}

\ 

We follow our paper [141] where some mathematical of the results of the paper
by Lazutkin and Pankratova (1975) were used\ for solution of concrete physical
problems. In particular, following our paper, let us consider the
Fuchsian-type equation given by
\begin{equation}
y^{^{\prime\prime}}+\frac{1}{2}\phi y=0, \tag{D.1}%
\end{equation}
where the potential $\phi$ is determined by the equation $\phi=[f]$ with
$f=y_{1}/y_{2}$ and $y_{1},y_{2}$ \ being two independent solutions of
Eq.(D.1) normalized by the requirement $y_{1}^{^{\prime}}y_{2}$ -$y_{2}%
^{\prime}$\ $y_{1}=1.$The symbol$\ [f]$ denotes the Schwarzian derivative of
$f$. Such a derivative is defined as follows
\begin{equation}
\lbrack f]=\frac{f^{\prime}f^{\prime\prime\prime}-\frac{3}{2}\left(
f^{\prime\prime}\right)  ^{2}}{\left(  f^{\prime}\right)  ^{2}}. \tag{D.2}%
\end{equation}
Consider Eq.(D.1) on the circle $S^{1}$ and consider some map of the circle
given by $F(t+1)=F(t)+1.$ Let $t=F(\xi)$ so that $y(t)=Y(\xi)\sqrt{F^{\prime
}(\xi)}$ leaves Eq.(D.1) form -invariant, i.e. in the form $Y^{\prime\prime
}+\frac{1}{2}\Phi Y=0$ with potential $\Phi$ being defined now as $\Phi
(\xi)=\phi(F(\xi))[F^{\prime}(\xi)]^{2}+[F(\xi)].$ Consider next the
infinitesimal transformation $F(\xi)=\xi+\delta\varphi(\xi)$ with $\delta$
being some small parameter and $\varphi(\xi)$ being some function to be
determined. Then, $\Phi(\xi+\delta\varphi(\xi))=\phi(\xi)+\delta(\hat
{T}\varphi)(\xi)+O(\delta^{2}).$ Here $(\hat{T}\varphi)(\xi)=\phi(\xi
)\varphi^{\prime}(\xi)+\frac{1}{2}\varphi^{\prime\prime\prime}(\xi
)+2\phi^{\prime}(\xi)\varphi(\xi).$ Next, we assume that the parameter
$\delta$ plays the same role as time. Then, we obtain
\begin{equation}
\lim_{t\rightarrow0}\frac{\Phi-\phi}{t}=\frac{\partial\phi}{\partial t}%
=\frac{1}{2}\varphi^{\prime\prime\prime}(\xi)+\phi(\xi)\varphi^{\prime}%
(\xi)+2\phi^{\prime}(\xi)\varphi(\xi) \tag{D.3}%
\end{equation}
Since thus far the perturbing function $\varphi(\xi)$ was left undetermined,
we can choose it now as $\varphi(\xi)=\phi(\xi).$ Then, we obtain the Korteweg
-de Vriez \ (KdV) equation%
\begin{equation}
\frac{\partial\phi}{\partial t}=\frac{1}{2}\phi^{\prime\prime\prime}%
(\xi)+3\phi(\xi)\phi^{\prime}(\xi) \tag{D.4}%
\end{equation}
determining the potential $\phi(\xi).$ For reasons which will be explained in
the text, it is sufficient to consider only the static case of KdV, i.e.%
\begin{equation}
\phi^{\prime\prime\prime}(\xi)+6\phi(\xi)\phi^{\prime}(\xi)=0. \tag{D.5}%
\end{equation}
We shall use this result as a reference for our main task of connecting the
Hill and the Neumann's problems. Using Eq.(54) we write%
\begin{equation}
u(\xi)=\phi(\xi)-<\dot{\xi},\dot{\xi}>. \tag{D.6}%
\end{equation}
Consider an auxiliary functional $\varphi(\xi)=<\xi,A^{-1}\xi>.$ Suppose that
$\varphi(\xi)=u(\xi).$ Then,
\begin{equation}
\frac{du}{dt}=2<\dot{\xi},A\xi>-2<\ddot{\xi},\dot{\xi}>. \tag{D.7}%
\end{equation}
But $<\ddot{\xi},\dot{\xi}>=0$ because of the normalization constraint
$<\xi,\xi>=1.$ Hence, $\dfrac{du}{dt}=2<\dot{\xi},A\xi>.$ Consider as well
$\dfrac{d\varphi}{dt}.$ By using Eq.s (54) it is straightforward to show that
$\dfrac{d\varphi}{dt}=2<\dot{\xi},A^{-1}\xi>.$ Because by assumption
$\varphi(\xi)=u(\xi)$ we have to demand that $<\dot{\xi},A^{-1}\xi>=<\dot{\xi
},A\xi>$ as well. If this is the case, consider furthermore
\begin{equation}
\dfrac{d^{2}u}{dt^{2}}=2<\ddot{\xi},A^{-1}\xi>+2<\dot{\xi},A^{-1}\dot{\xi}>
\tag{D.8}%
\end{equation}
Using Eq.s(54) once again we obtain%
\begin{equation}
\dfrac{d^{2}u}{dt^{2}}=-2+2u\varphi+2<\dot{\xi},A^{-1}\dot{\xi}>. \tag{D.9}%
\end{equation}
Finally, consider as well $\dfrac{d^{3}u}{dt^{3}}.$ Using Eq.(D.9) as well as
Eq.(54) and (D.7) we obtain,%
\begin{equation}
\dfrac{d^{3}u}{dt^{3}}=2\frac{du}{dt}\varphi+4u\frac{du}{dt}=6u\frac{du}{dt}
\tag{D.10}%
\end{equation}
By noticing that in Eq.(D.5) we can always make a rescaling $\phi
(\xi)\rightarrow\lambda\phi(\xi)$ we always can choose $\lambda=-1.$Therefore
Eq.s (D.5) and (D.10) coincide. This establishes the correspondence between
the Neumann and Hill-type problems.

QED

\bigskip

\bigskip

{\Large References}

\bigskip

\bigskip

1.Born, M.: The Mechanics of Atom. Frederic Ungar Publishing Co,

\ \ \ New York, (reproduced from 1924 lecture notes) (1960)

2.Porter, M., Cvitanovi\v{c}, P.: Ground control to Niels Bohr: Exploring

\ \ \ outer space with atomic physics. AMS Notices \ \textbf{52},1020-1025 (2005)

3.Marsden, J., Ross, S.: New methods in celestial mechanics and mission

\ \ \ design. AMS Bulletin \textbf{43}, 43-73 (2006)

4.Convay, B., Chilan, C., Wall, B.: Evolutionary principles applied to

\ \ \ mission planning problems.

\ \ \ Celest. Mech. Dynam.Astron. \ \textbf{97}, 73-86 (2007)

5. Pauli, W., Born, M.: About quantization of perturbed mechanical

\ \ \ \ system (in German). Z.Phys. \textbf{10}, 137-158 (1922)

6. Baly, E.: Spectroscopy. Longmans,Green and Co., New York (1905)

7. Bohr, N.: The spectra of Helium and Hydrogen.

\ \ \ \ Nature \textbf{92}, 231-232 (1913)

8. Ingle, J., Couch, S.: Spectrochemical Analysis. Prentice Hall (1988)

9. Tanner, G., Richter, K., Rost, J-M.: The theory of two-electron atoms:

\ \ \ \ between ground state and complete fragmentation.

\ \ \ \ Rev. Mod.Phys. \textbf{72} (2) 497-544 (2000)

10. Svidzinsky, A., Scully, M., Herschbach, D.: Bohr's 1913 molecular model

\ \ \ \ \ revisited. PNAS \textbf{102} (34), 11985-11988 (2005)

11. Svidzinsky, A., Scully, M., Herschbach, D.: Simple and surprisingly

\ \ \ \ \ accurate approach to the chemical bond obtained from dimensional

\ \ \ \ \ scaling. PRL \ \textbf{95 }080401-1-080401-4 (2005)

12. Murawski, R., Svidzinsky, A.: Quantum number dimensional scaling

\ \ \ \ \ analysis for excited states of multielectron atoms.

\ \ \ \ \ arXiv:physics/0610056

13. Chen, G., Hsu, S-B., Kim, M., Zhou, J.:Mathematical analysis of a Bohr

\ \ \ \ \ atom model. JMP \textbf{47}, 022107-1-022107-22 (2006)

14. Molchanov, A.: The resonant structure of the Solar System.

\ \ \ \ \ The law of planetary distances. Icarus \textbf{8}, 203-215 (1968)

15. Ferraz-Mello, S., Michtchenko, T., Beauge, C.: Regular motions

\ \ \ \ \ in extrasolar planetary systems. In Chaotic Worlds: From Order

\ \ \ \ to Disorder in Gravitational N-Body System.

\ \ \ \ \ Springer, Heidelberg (2006)

16. Ferraz-Mello, S., Michtchenko, T., Beauge, C., Callegari, N.:

\ \ \ \ \ Extrasolar planetary systems. In Chaos and Stability in Extrasolar

\ \ \ \ \ Planetary Systems. Springer, Heidelberg (2005)

17. Beletsky, V.: Essays on the Motion of Celestial Bodies

\ \ \ \ \ Birkh\"{a}user, Boston (2001)

18. Arnol'd, V., Kozlov,V., Neishtadt, A.: Mathematical Aspects

\ \ \ \ \ of Classical and Celestial Mechanics. Springer, Heidelberg (2006)

19. \ Henon, M.: A comment on "The resonant structure of the Solar System "

\ \ \ \ \ \ by A.M.Molchanov. Icarus \textbf{11}, 93-94 (1969)

20. \ Backus, G.: Critique of \ "The resonant structure of the Solar System"

\ \ \ \ \ \ by A.M.Molchanov. Icarus \textbf{11}, 88-92 (1969)

21. \ Molchanov, A.: Resonances in complex systems: a reply to critiques.

\ \ \ \ \ \ Icarus \textbf{11}, 95-103 (1969)

22. \ Molchanov, A.: The reality of resonances in the Solar System.

\ \ \ \ \ \ Icarus \textbf{11}, 104-110 (1969)

23. \ Brin, J.: On the stability of the planetary system.

\ \ \ \ \ \ Astron.\&Astrophys. \textbf{24}, 283-293 (1973)

24. \ Patterson, C.: Resonance capture and the evolution of the planets.

\ \ \ \ \ \ Icarus \textbf{70}, 319-333 (1987)

25. \ Goldreich, P. An explanation of the frequent occurence of

\ \ \ \ \ \ commensurable mean motions in the Solar System.

\ \ \ \ \ \ Mon. Not.R. Astr.Soc. \textbf{130}, 159-181 (1965)

26. \ Murray, C., Dermott, S.: Solar System Dynamics.

\ \ \ \ \ \ Cambridge University Press, Cambridge (1999)

27. \ \ Heisenberg,W.: Quantum-theoretical re-interpretation of kinematic

\ \ \ \ \ \ \ and mechanical relations.(in German) Z.Phys. \textbf{33},
879-893 (1925)

28. \ \ Knutson, A., Tao,\ T.: Honeycombs and sums of \ Hermitian matrices.

\ \ \ \ \ \ \ \ AMS Notices \ \textbf{48},175-186 (2001)

29. \ \ Kholodenko, A.: Heisenberg honeycombs solve Veneziano

\ \ \ \ \ \ \ puzzle. International Mathematical Forum (2008) in press,

\ \ \ \ \ \ \ arxiv: hep-th/0608117

30. \ \ Knutson, A., Tao, T.: The honeycomb model of GL$_{n}(\mathbf{C})$ tensor

\ \ \ \ \ \ \ products. I. Proof of the saturation conjecture.

\ \ \ \ \ \ \ J.AMS \textbf{12},1055-1090\ (1999)

31. \ \ Knutson, A., Tao,T., Woodward,C.: The honeycomb model

\ \ \ \ \ \ \ of GL$_{n}(\mathbf{C})$ tensor products. II. Puzzles determine
facets of

\ \ \ \ \ \ \ the Littlewood-Richardson cone. J.AMS \textbf{17},19-48 (2004)

32. \ \ Tao, T.: [www.math.ucla.edu/\symbol{126}tao/java/Honeycomb.html] (2001)

33. \ \ Dirac, P.: The fundamental equations of quantum mechanics.

\ \ \ \ \ \ \ Proc.Roy.Soc. A \textbf{109}, 642-653 (1926)

34. \ \ Dirac, P.: Principles of Quantum Mechanics. Clarendon Press,

\ \ \ \ \ \ \ Oxford, UK (1958)

35. \ \ Kholodenko, A.: New strings for old Veneziano amplitudes.

\ \ \ \ \ \ \ II Group-theoretic treatment. J.Geom.Phys. \textbf{56},1387-1432 (2006)

36. \ \ Kirillov, A.: Elements of the Theory of Representations.

\ \ \ \ \ \ \ \ Springer-Verlag, Heidelberg (1976)

37. \ \ \ Kargapolov, M., Merzlyakov, Ju.: Fundamentals of the Theory

\ \ \ \ \ \ \ \ of Groups. Springer-Verlag, Heidelberg (1979)

38. \ \ \ Vershik, A.: Asymptotic Combinatorics with Applications

\ \ \ \ \ \ \ \ to Mathematical Physics. Springer-Verlag, Heidelberg (2003)

39. \ \ \ Fulton,W.:Young Tableaux.

\ \ \ \ \ \ \ \ Cambridge University Press, Cambridge (1997)

40. \ \ \ Borel, A.: Linear Algebraic Groups.

\ \ \ \ \ \ \ \ Springer-Verlag, Heidelberg (1991)

41. \ \ \ Di Francesco, P., Mathieu, P., Senechal, D.: Conformal Field

\ \ \ \ \ \ \ \ Theory. Springer-Verlag, Heidelberg (1997)

42. \ \ \ Vershik, A.,Okounkov, A.: A new approach to the representation

\ \ \ \ \ \ \ theory of the symmetric groups II.

\ \ \ \ \ \ J.Math.Sciences \textbf{131}, 5471-5494 (2005)

43. \ \ Macdonald, I.: Symmetric Functions and Orthogonal Polynomials

\ \ \ \ \ \ \ AMS Publishers, Providence, RI (1998)

44. \ \ Dittrich, W., Reuter, M.: Classical and Quantum

\ \ \ \ \ \ \ Dynamics. Springer-Verlag, Heidelberg (1992)

45. \ \ Arnol'd, V.: Mathematical Methods of Classical

\ \ \ \ \ \ \ \ Mechanics. Nauka, Moscow \ (1974)

46. \ \ \ Kuznetsov, V.: Orthogonal polynomials and separation of variables.

\ \ \ \ \ \ \ \ LNM \ \textbf{1883} , 229-254 (2006)

47. \ \ \ Baik, J., Kriecherbauer, T., McLaughlin, K., Miller, P.: Discrete

\ \ \ \ \ \ \ \ Orthogonal Polynomials: Asymptotics and Applications.

\ \ \ \ \ \ \ \ Princeton U.Press, Princeton (2007)

48. \ \ \ Orlik, P., Terrao, H.: Arrangements and Hypergeometric

\ \ \ \ \ \ \ \ Functions. MSJ Publications, Tokyo (2001)

49. \ \ \ Flugge, S.: Practical Quantum Mechanics. Springer-Verlag,

\ \ \ \ \ \ \ \ Heidelberg (1971)

50. \ \ \ Vassiliev, V.:Applied Picard-Lefschetz Theory. AMS Publishers,

\ \ \ \ \ \ \ \ \ Providence, RI (2002)

51. \ \ \ \ Lascoux, A., Sch\"{u}tzenberger, M.:Symmetry and flag manifolds.

\ \ \ \ \ \ \ \ \ LNM \ \textbf{996}, 118- 144\ (1983)

52. \ \ \ \ Kirillov Jr., A.: \ Lectures on affine Hecke algebras and Macdonald

\ \ \ \ \ \ \ \ \ conjectures. AMS Bulletin \ \textbf{34}, 251-292 (1997)

53. \ \ \ \ \ Adin, R., Postnikov, A., Roichman, Y.: Hecke algebra actions on the

\ \ \ \ \ \ \ \ \ coinvariant algebra. J.of Algebra \textbf{233}, 594-613 (2000)

54. \ \ \ \ Kassel, K.: Quantum Groups. Springer-Verlag, Heidelberg (1995)

55. \ \ \ \ Belavin, A.,\ Drinfel'd, V.: Solutions of the classical Yang-Baxter

\ \ \ \ \ \ \ \ \ equations\ for simple Lie algebras.

\ \ \ \ \ \ \ \ \ Funct. Anal. Appl.(in Russian) \textbf{16}, 1-29 (1982)

56. \ \ \ \ Etingoff, P., Frenkel, I., Kirillov Jr., A.: Lectures on Representation

\ \ \ \ \ \ \ \ \ Theory and Knizhnik-Zamolodchikov Equations.

\ \ \ \ \ \ \ \ \ AMS Publishers, Providence, RI (1998)

57. \ \ \ \ Tanner, G., Richter, K., Rost, J-M.: The theory of two-electron

\ \ \ \ \ \ \ \ \ atoms: between ground state and complete fragmentation.

\ \ \ \ \ \ \ \ \ Rev. Mod.Phys. \textbf{72} (2). 497-544 (2000)

58. \ \ \ \ Cvitanovic$^{\prime},P.:$ Classical and Quantum Chaos: A Cyclist Treatise.

\ \ \ \ \ \ \ \ \ \ http:/www.nbi.dk/ChaosBook/ (1998)

59. \ \ \ \ \ Draeger, M., Handke, G., Ihra, W., Friedrich, H.: One and

\ \ \ \ \ \ \ \ \ \ two-electron excitations of helium in the s-wave model.

\ \ \ \ \ \ \ \ \ \ Phys.Rev.A \ \textbf{50}, 3793-3808 (1994)

60. \ \ \ \ \ Howard, I., March, N.: Towards a differential equation for the

\ \ \ \ \ \ \ \ \ \ nonrelativistic ground-state electron density of the
He-like sequence

\ \ \ \ \ \ \ \ \ \ of atomic ions. Phys.Rev A \textbf{71, }042501-1-042501-5 (2005)

61. \ \ \ \ \ Ostrovsky, V., Prudov, N.: Planetary atom states: adiabatic

\ \ \ \ \ \ \ \ \ \ invariant theory. J.Phys.B \textbf{28}, 4435-4445 (1995)

62. \ \ \ \ \ Ferraz-Mello, S.: Canonical Perturbation Theories.

\ \ \ \ \ \ \ \ \ Springer, Heidelberg (2007)

63. \ \ \ \ Pars, L.: Analytical Dynamics. Heinemann, London (1968)

64. \ \ \ \ Dirac, P.: Generalized Hamiltonian dynamics.

\ \ \ \ \ \ \ \ \ Canadian J.Math. \textbf{2}, 129-148(1950)

65. \ \ \ \ Bander, M., Itzykson, C.: Group theory and the hydrogen atom.

\ \ \ \ \ \ \ \ \ Rev. Mod. Phys. \textbf{38}, 330-358 (1966)

66. \ \ \ \ Kac, M.: Can one hear the shape of a drum?

\ \ \ \ \ \ \ \ \ Am.Math.Monthly \textbf{73}, 1-23 (1966)

67. \ \ \ \ Dhar, A., Rao, D., Shankar, U., Sridhar, S.: Isospectrality in chaotic

\ \ \ \ \ \ \ \ \ billiards. Phys.Rev.E \textbf{68}, 026208-1-0262081-5 (2003)

68. \ \ \ Landau, L., Lifshitz, E.: Classical Mechanics.

\ \ \ \ \ \ \ \ Pergamon, London (1960)

69. \ \ \ Jauch, J., Hill, E.: On the problem of degeneracy in quantum

\ \ \ \ \ \ \ \ mechanics. Phys.Rev. \textbf{57}, 641-645 (1940)

70. \ \ \ Wybourne, B.: Classical Groups for Physicists. John Willey \& Sons,

\ \ \ \ \ \ \ \ New York (1974)

71. \ \ \ Natanzon, G.: General properties of potentials for which the

\ \ \ \ \ \ \ \ Schr\"{o}dinger equation is solvable by means of hypergeometric

\ \ \ \ \ \ \ \ functions.\ Theor..Math.Phys. \textbf{38}, 219-229 (1979)

72. \ \ \ Levai, G.: Solvable potentials associated with su(1,1) algebras:

\ \ \ \ \ \ \ \ a systematic study. J.Phys.A \textbf{27}, 3809-3828 (1994)

73. \ \ \ Cordero, P., Holman, S., Furlan, P., Ghirardi, G.: Algebraic

\ \ \ \ \ \ \ \ treatment of \ nonrelativistic and relativistic quantum equations

\ \ \ \ \ \ \ \ and its relation to the theory of differential equations.

\ \ \ \ \ \ \ \ Il Nuovo Cim. \textbf{3A}, 807-821 (1971)

74. \ \ \ Bargmann, V.: Irreducible unitary representations of the

\ \ \ \ \ \ \ \ Lorentz group.Ann.Math. \textbf{48}, 568-640 (1947)

75. \ \ \ Barut, A., Fronsdal, C.: On non compact groups.

\ \ \ \ \ \ \ \ II Representations of the 2+1 Lorentz group.

\ \ \ \ \ \ \ \ Proc.Roy.Soc.London \textbf{A} \textbf{287}, 532-548 (1965)

76. \ \ \ Cooper, F., Ginoccio, J., Khare, A.: Relationship between the

\ \ \ \ \ \ \ \ supersymmetry and solvable potentials.

\ \ \ \ \ \ \ \ Phys.Rev.\textbf{D 36}, 2458-2473 (1987)

77. \ \ \ Junker, G., Roy, P.:\ Conditionally exactly solvable potentials: a

\ \ \ \ \ \ \ \ \ supersymmetric construction method. Ann.Phys.\textbf{270},
155-177 (1998)

78. \ \ \ Boyer, T.: Random electrodynamics. Phys.Rev.D \textbf{11}, 790-808 (1975)

79. \ \ \ Puthoff, H.: Ground state of hydrogen as a zero point fluctuation

\ \ \ \ \ \ \ \ determined state.\ Phys.Rev. \textbf{D 35}, 3266-3269 (1987)

80. \ \ \ Goldstein, H., Poole,C., Safko, J.: Classical Mechanics.

\ \ \ \ \ \ \ \ \ Addison-Wesley, New York (2002)

81. \ \ \ Landau, L., Lifshitz E.: The Classical Theory of Fields.

\ \ \ \ \ \ \ \ Pergamon, London (1975)

82. \ \ \ Einstein, A.: Foundations of the general theory of relativity

\ \ \ \ \ \ \ \ (in German). Ann.Phys. \textbf{49}, 769-822 (1916)

83. \ \ \ Misner, C., Thorne, K., Wheeler, J.: Gravitation.

\ \ \ \ \ \ \ \ Freeman \& Co, New York (1973)

84. \ \ \ Moser, J., Zehnder, E.: Notes on Dynamical Systems.

\ \ \ \ \ \ \ \ AMS Publishers, Providence, RI (2005)

85. \ \ \ Charlier, C.: Die Mechanik des Himmels.

\ \ \ \ \ \ \ \ Walter de Gruyter, Berlin (1927)

86. \ \ \ Fejoz, J.: Demonstration 'du theoreme d'Arnold' sur la
stabilite$^{\prime}$

\ \ \ \ \ \ \ \ du systeme planetaire (d'apres Herman).

\ \ \ \ \ \ \ \ Erg.Th.\&Dyn.Syst. \textbf{24}, 1521-1582 (2004)

87. \ \ \ Biasco, L., Cherchia, L., Valdinoci, E.: N-dimensional elliptic

\ \ \ \ \ \ \ \ invariant tori for planar (N+1) body problem.

\ \ \ \ \ \ \ \ SIAM J.Math.Anal. \textbf{37}, 1580-1598 (2006)

88. \ \ \ Poincare$^{\prime}$, H.: Les Methodes Nouvelles de la

\ \ \ \ \ \ \ \ \ Mechanique Celeste. Gauthier-Villars, Paris (1892-1899)

89. \ \ \ Einstein, A., Infeld, L., Hoffmann, B.:The gravitational equations

\ \ \ \ \ \ \ \ and the problem of motion. Ann.Math. \textbf{39}, 65-100 (1938)

90. \ \ \ Robertson, H.: Note on the paper by Hoffman, Infeld and Einstein.

\ \ \ \ \ \ \ \ \ Ann.Math. \textbf{39}, 101-104 (1938)

91. \ \ \ \ Nieto, M.: The Titius -Bode Law of Planetary Distances:

\ \ \ \ \ \ \ \ \ Its History and Theory. Pergamon Press, London (1972)

92. \ \ \ \ Celletti, A., Perozzi, E.: Celestial Mechanics.

\ \ \ \ \ \ \ \ \ The Waltz of the Planets. Springer, Heidelberg (2007)

93. \ \ \ \ Neslu\v{s}an, L.: The significance of the Titius-Bode law and

\ \ \ \ \ \ \ \ \ the peculliar location of the Earth's orbit.

\ \ \ \ \ \ \ \ \ Mon. Not.R.Astr.Soc. \textbf{351}, 133-136 (2004)

94. \ \ \ \ Lynch, P.: On the significance of the Titius-Bode law of the

\ \ \ \ \ \ \ \ \ distribution of the planets.

\ \ \ \ \ \ \ \ \ Mon.Not. R.Astr.Soc. \textbf{341}, 1174-1178 (2003)

95. \ \ \ \ Dermott, S.: On the origin of commensurabilities in the

\ \ \ \ \ \ \ \ \ Solar System II. Mon.Not.R.Astr.Soc. \textbf{141}, 363-376 (1968)

96. \ \ \ \ Hayes,W., Tremaine, S.: Fitting selected random planetary systems

\ \ \ \ \ \ \ \ \ to Titius-Bode laws. Icarus \textbf{135}, 549-557 (1998)

97. \ \ \ \ Landau, L., Lifshitz, E.: Quantum Mechanics.

\ \ \ \ \ \ \ \ \ Pergamon, London (1962)

98. \ \ \ \ Chebotarev, G.: Analytical and Numerical Methods of Celestial

\ \ \ \ \ \ \ \ \ Mechanics. Elsevier, Amsterdam (1968)

99. \ \ \ \ Milczewski, J, Farrelly, D., Uzer, T.: 1/r dynamics in external
fields :

\ \ \ \ \ \ \ \ \ 2D or 3D? PRL \textbf{78}, 2349-2352 (1997)

100. \ \ Jaffe, C., Ross, S., Lo, M., Marsden, J., Farrelly, D.,Uzer, T.:

\ \ \ \ \ \ \ \ \ Statistical theory of asteroid escape rates.

\ \ \ \ \ \ \ \ \ PRL \textbf{89}, 011101-1-011101-4 (2002)

101. \ \ Micha, D., Burghardt, I.: Quantum Dynamics of Complex

\ \ \ \ \ \ \ \ \ Molecular Systems. Springer, Heidelberg (2007)

102. \ \ Szebehely,V.: Theory of Orbits. Academic Press, New York (1967)

103. \ \ Avron, J., Simon, B.: Almost periodic Hill's equation and the rings

\ \ \ \ \ \ \ \ \ of Saturn. PRL \textbf{46}, 1166-1168 (1981)

104. \ \ Brouwer, D., Clemence, G.: Methods of Celestial Mechanics.

\ \ \ \ \ \ \ \ \ Academic Press. New York (1961)

105. \ Ince, E.: Ordinary Differential Equations.

\ \ \ \ \ \ \ \ Dover Publishers, New York (1926)

106. \ Magnus,W.,Winkler, S.: Hill's Equation. Interscience Publishers.

\ \ \ \ \ \ \ \ New York (1966)

107. \ Mc Kean, H., \ Moerbeke, P.: The spectrum of Hill's equation.

\ \ \ \ \ \ \ \ Inv.Math. \textbf{30}, 217- 274 (1975)

108. \ Trubowitz, E.: The inverse problem for periodic potentials

\ \ \ \ \ \ \ \ Comm. Pure Appl. Math. \textbf{30}, 321-337 (1977)

109. \ Moser, J.: \ Various aspects of integrable Hamiltonian systems.

\ \ \ \ \ \ \ In: Dynamical Systems, pp.233-290 , Birkh\"{a}user, Boston (1980)

110. \ Hochstadt, H.: Functiontheoretic properties of the discriminant

\ \ \ \ \ \ \ \ of Hill's equation. Math.Zeitt. \textbf{82}, 237-242 (1963)

111. \ Esposito, L.: Planetary Rings. Cambridge University Press.

\ \ \ \ \ \ \ \ Cambridge UK (2006)

112. \ Kirillov, A.: Infinite dimensional Lie groups: their orbits,

\ \ \ \ \ \ \ \ invariants and \ representations. LNM\ \ \textbf{970}, 101-123 (1982)

113. \ Vilenkin, N.: Special Functions and Theory of Group Representations

\ \ \ \ \ \ \ \ (in Russian) Nauka, Moscow (1991)

114. \ Veselov, A., Dullin, H., Richter, P., Waalkens, H.: Actions of

\ \ \ \ \ \ \ \ the Neumann systems via Picard-Fuchs equations.

\ \ \ \ \ \ \ \ Physica \textbf{D} \textbf{155}, 159-183 (2001)

115. \ Utiyama, R.: Invariant theoretical interpretation of interaction.

\ \ \ \ \ \ \ \ Phys.Rev.\textbf{101}, 1597- (1956)

116. \ Dirac, P.: The electron wave equation in de-Sitter space.

\ \ \ \ \ \ \ \ Ann.Math. \textbf{36}, 657-669 (1935)

117. \ Aldrovandi, R., Beltran Almeida, J., Pereira, J.: Cosmological term

\ \ \ \ \ \ \ \ and \ fundamental physics [arXiv: gr-qc/0405104] (2004)

118. \ Bros, J., Epstein, H., Moschella, U.: Lifetime of a massive particle

\ \ \ \ \ \ \ \ in a de Sitter universe. [arxiv: hep-th/0612184] (2006)

119. \ Peebles, P.Ratra, B.: The cosmological constant and dark energy.

\ \ \ \ \ \ \ \ Rev.Mod Phys.\textbf{75}, 559-606 (2003)

120. \ Copeland, E., Sami, M .Tsujikawa, S.: Dynamics of dark energy

\ \ \ \ \ \ \ \ [arxiv: hep-th/0603057] (2006)

121. \ Kay, S., Pearce, F., Frenk, C., Jenkins, A.: Including star formation

\ \ \ \ \ \ \ \ and supernova feedback within cosmological simulations

\ \ \ \ \ \ \ \ of galaxy formation Mon.Not.R. Astron. Soc. \textbf{330},
113-128 (2002)

122. \ \ De Blok, W., McGauh, S, Bosma, A.. Rubin,V.: Mass density profiles

\ \ \ \ \ \ \ \ of low surface brightness galaxies.

\ \ \ \ \ \ \ \ Astrophys. J.Lett. \textbf{552}, L 23-26(2001)

123. \ \ Kholodenko, A.: Towards physically motivated proofs of the
Poincare$^{\prime}$

\ \ \ \ \ \ \ \ \ and the geometrization conjectures. J.Geom.Phys.
\textbf{58}, 259-290 (2008)

124. \ Willmore, T.: Riemannian Geometry. Clarendon Press, Oxford (1993)

125. \ Hawking, S., Ellis, G.: The Large Scale Structure of Space-Time.

\ \ \ \ \ \ \ \ Cambridge University Press, Cambridge, UK (1973)

126. \ Graner, F., Dubrulle, B.: Titius-Bode law in the Solar System I.

\ \ \ \ \ \ \ \ Scale\ invariance explains everything.

\ \ \ \ \ \ \ \ Astron.Astrophys. \textbf{282}, 262-268 (1994)

127. \ Petrov, A.: Einstein Spaces. Pergamon, Oxford (1969)

128. \ Weinberg, S.: Gravitation and Cosmology. John Willey \& Sons,

\ \ \ \ \ \ \ \ New York (1972)

129. \ Andersson, L, Barbot, T., Beguin, F., Zeghib, A.: Cosmological

\ \ \ \ \ \ \ \ time versus CMS time II: the de Sitter and anti-de Sitter cases.

\ \ \ \ \ \ \ \ [arxiv: math/0701452] (2007)

130. \ Kholodenko, A.: Use of meanders and train tracks for description

\ \ \ \ \ \ \ \ of defects and textures in liquid crystals and 2+1 gravity.

\ \ \ \ \ \ \ J.Geom.Phys. \textbf{33}, 23-58 (2000)

131. \ Kholodenko, A : Use of quadratic differentials for description of defects

\ \ \ \ \ \ \ \ and textures in 2+1 gravity. J.Geom.Phys, \textbf{33}, 59-102 (2000)

132. Kholodenko, A : Statistical mechanics of 2+1 gravity from Riemann

\ \ \ \ \ \ zeta function and Alexander polynomial: exact results.

\ \ \ \ \ \ \ J.Geom.Phys. \textbf{38}, 81-139 (2001)

133. 't Hooft, G.:A locally finite model for gravity.

\ \ \ \ \ \ \ Found.Phys.\textbf{38},733-757 (2008)

134. Thurston, W.: Three-dimensional manifolds, Kleinian groups

\ \ \ \ \ \ \ and hyperbolic geometry. AMS Bull. \textbf{6}, 357-381 (1982)

135. Deser, S., Jackiw, R., t'Hooft, G.: Three-dimensional Einstein

\ \ \ \ \ \ \ gravity: dynamics in flat space. Ann.Phys.\textbf{152}, 220-235 (1984)

136. Nikolaev, I.: Foliations on Surfaces.

\ \ \ \ \ \ \ Springer-Verlag, Heidelberg (2001)

137. Hehl, F., Obukhov, Y.: Foundations of Classical Electrodynamics.

\ \ \ \ \ \ \ Birk\"{a}user, Boston (2003)

138. \ Moerdijk, I., Mr\v{c}un, J. : Introduction to Foliations and Lie Groupoids.

\ \ \ \ \ \ \ Cambridge U.Press, Cambridge (2003)

138. Witten, E.: Three-dimensional gravity revisited.

\ \ \ \ \ \ \ arXiv: 0706.3359 (2007)

139. \ Kholodenko, A.: Veneziano amplitudes, spin chains and

\ \ \ \ \ \ \ \ Abelian reduction of QCD. arXiv: 0810.0250

140. \ Davydov, A.: Quantum mechanics, Pergamon, London (1965)

141. \ Kholodenko, A.: Kontsevich-Witten model from 2+1 gravity:

\ \ \ \ \ \ \ \ new exact \ combinatorial solution. J.Geom.Phys. \textbf{43},
45-91 (2002)

\bigskip

\bigskip

\bigskip

\bigskip

\bigskip

\bigskip

\bigskip

\bigskip
\end{document}